\documentclass[twocolumn]{aastex7}

\usepackage{amsmath,amssymb,amstext,fp,mathtools,xspace}

\newcommand{\Cov}{\operatorname{Cov}}

 \newcommand{\f}[1]{\mathrm{\uppercase{f{#1}}}}
 \let\F=\f

 \let\HST=\hst

 \let\JWST=\jwst

 \let\roman=\rst
 \let\Roman=\rst
\newcommand{\Euclid}{\emph{Euclid}\xspace}

 \makeatletter
 \renewcommand{\sectionautorefname}{\S\@gobble}
 \renewcommand{\subsectionautorefname}{\S\@gobble}
 \renewcommand{\subsubsectionautorefname}{\S\@gobble}
 \makeatother

\graphicspath{{./}{}}

\begin{document}

\title{NIR filter transformations across the HR diagram: JWST, Roman, and Euclid}
\shorttitle{NIR filter transformations}
\shortauthors{M.~J.~Durbin et al.}

\correspondingauthor{Meredith Durbin}
\email{meredith.durbin@berkeley.edu}

\author[0000-0001-7531-9815]{M.~J.~Durbin}
    \affiliation{Department of Astronomy, University of California, Berkeley, Berkeley, CA, 94720, USA}
\email{meredith.durbin@berkeley.edu}

\author[0000-0002-1691-8217]{R.~L.~Beaton}
    \affil{Space Telescope Science Institute, Baltimore, MD, 21218, USA}
    \affil{Department of Physics and Astronomy, Johns Hopkins University, Baltimore, MD 21218, USA}
\email{rbeaton@stsci.edu}

\author[0000-0002-0048-2586]{A.~J.~Monson}
    \affiliation{Department of Astronomy/Steward Observatory, University of Arizona, 933 North Cherry Avenue, Tucson, AZ 85721, USA}
\email{andymonson@arizona.edu}

\author[0000-0002-3749-4978]{B.~Swidler}
    \affiliation{AFWERX, Air Force Research Laboratory, Wright-Patterson AFB, Dayton, Ohio, USA}
\email{biancaswidler@gmail.com}

\begin{abstract}

We present new color transformations between select near-infrared filters on \JWST/NIRCam, \Euclid/NISP, \Roman/WFI, \HST, and ground-based $izY$+$IJHK_S$, for a total of 105 unique filter combinations.
Additionally, we apply these transformations to predict the color-magnitude relation of the tip of the red giant branch as seen with \JWST, \Euclid, and \Roman based on theoretical results for \HST and 2MASS filters; for \JWST we find good agreement with empirical results in the literature. 
We also find typical residual dispersion around these transformations of 0.01~mag for Cepheid and RR Lyrae variables and RGB stars, but up to 0.1 mag for O- and C-rich TP-AGB stars. 

\end{abstract}

\keywords{Calibration (2179), infrared astronomy (786), photometric systems (1233), stellar colors (1590), stellar populations (1622)}

\section{Introduction} \label{sec:intro}

Near-infrared (NIR) spectrophotometry of stars is becoming increasingly crucial for a variety of astrophysical studies. Recent and upcoming missions that operate predominantly in the NIR, such as \added{the \emph{James Webb Space Telescope} \citep[\JWST,][]{2023PASP..135d8001R}, the \Euclid mission \citep{2025A&A...697A...1E}, and the \emph{Nancy Grace Roman Space Telescope}} \citep{2019arXiv190205569A}, are expected to observe many more stars with greater accuracy and at farther distances than any previous missions by the end of their lifetimes.
\Euclid and \Roman in particular are survey telescopes covering large areas of the sky, and as such are designed to accommodate a wide range of scientific use cases in addition to their key projects.

\begin{figure}[htb]
    \centering
    \includegraphics[width=\columnwidth]{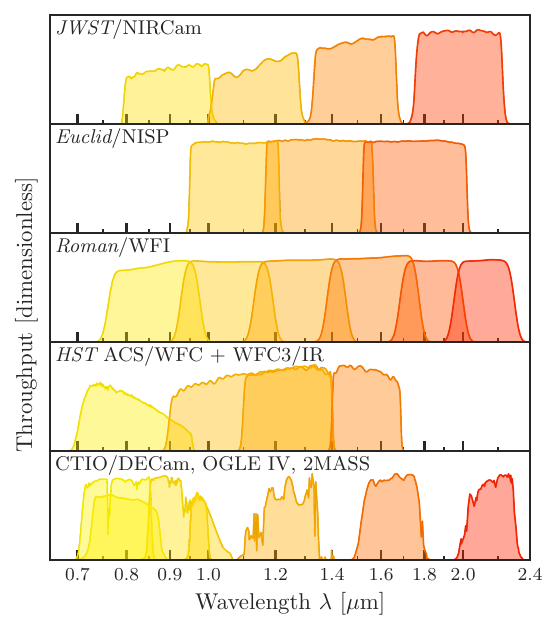}
    \caption{Throughput curves for all filters considered in this work.}
    \label{fig:throughputs}
\end{figure}

Unlike their ground-based counterparts, spaceborne observatories are not constrained by atmospheric transmission windows, and may access the full range of NIR wavelengths.
This can result in substantially greater variations than expected between bandpasses that nominally correspond to the same ground-based filter.
\Euclid and \Roman specifically are optimized for continuous wavelength coverage, in part for \added{spectral energy distribution} (SED) reconstruction and photometric redshift measurement.
This is illustrated in \autoref{fig:throughputs}, which compares throughput curves for all the facilities we consider in this paper. 

This paper is a companion to \citet{2023AJ....166..236D}, which focused on transformations between \added{the infrared channel of \HST's Wide Field Camera 3} (WFC3/IR) and the \added{Two Micron All Sky Survey} (2MASS).
Here we extend the methods and data presented there to newer and upcoming missions, for which interest in joint calibration is high \citep[e.g.][]{2020arXiv200810663C, 2021arXiv210401199R, 2021EPSC...15..298B}. %
\autoref{sec:obs} gives an overview of our data and methods.
We present the resulting transformations in \autoref{sec:res}, and use them to make predictions for the tip of the red giant branch in \JWST, \Roman, and \Euclid filters.
We consider optimal filter combinations for various stellar populations applications in \autoref{sec:disc},
and discuss conclusions and future work in \autoref{sec:conc}.

\section{Data and Methods} \label{sec:obs}

Our input datasets and synthetic photometry technique are described in detail by \citet{2023AJ....166..236D}. 
In this section we provide a brief summary of the process, and discuss minor updates where applicable.

\subsection{Data} \label{ssec:data}

We combine data products from four spectral libraries to maximize coverage of the full HR diagram at the relevant wavelengths.
These are the \HST flux standard library CALSPEC \citep[][and references therein]{1990AJ.....99.1243T, 2001AJ....122.2118B, 2007ASPC..364..315B, 2008AJ....136.1171B, 2014PASP..126..711B, 2019AJ....157..229B}; the NASA Infrared Telescope Facility original and extended libraries (IRTF and EIRTF respectively, \citealt{2009ApJS..185..289R} and \citealt{2017ApJS..230...23V}); and the X-Shooter Spectral Library third data release \citep[XSL,][]{2014AA...565A.117C, 2020AA...634A.133G, 2022AA...660A..34V}.
\added{The total sample comprises 1269 spectra of 1061 stars, with apparent magnitudes $-4.1 \le m_{K_S} \le 17.4$, absolute magnitudes $-12.5 \le M_{K_S} \le 13.5$, and colors $-0.3 \le J-K_S \le 4.1$.}

We deredden all library spectra with line-of-sight reddenings adopted from the 3D dust map of \citet[][updated by \citealt{2022AA...664A.174V}]{2022AA...661A.147L}
for Galactic stars, with the majority of distances from \citet{2021AJ....161..147B}.
For the small subset of library stars that are Magellanic Cloud members, we use the 2D reddening map of \citet{2021ApJS..252...23S}.

Throughputs for \HST, 2MASS, and \Roman/WFI are available via the \HST calibration reference data system (CRDS)\footnote{\url{https://hst-crds.stsci.edu/}} under the context \texttt{hst\_synphot\_0067.imap}.
\JWST throughputs are described by and available through the JWST User Documentation.\footnote{\url{https://jwst-docs.stsci.edu/jwst-near-infrared-camera/nircam-instrumentation/nircam-filters}}
The \Euclid/NISP filter curves are described by \citet{2022A&A...662A..92E} and available at the \Euclid Mission Reference Data website.\footnote{\url{https://doi.org/10.5270/esa-kx8w57c}}
(Note that we adopt only the detector-averaged \JWST/NIRCam filter curves and do not take into account chip-dependent throughput differences, nor do we account for variations in filter cutoff wavelengths over the \Euclid/NISP and \Roman/WFI fields of view.)
The OGLE-IV $I$ filter, a close approximation of the standard Cousins $I$, is described by \citet{2015AcA....65....1U} and available on the OGLE project website,\footnote{\url{https://ogle.astrouw.edu.pl/main/OGLEIV/mosaic.html}} and the CTIO/DECam filters are described by \citet{2018ApJS..239...18A} and available via NOIRLab.\footnote{\url{https://noirlab.edu/science/programs/ctio/filters/Dark-Energy-Camera}}

Our choice of CTIO/DECam to represent ground-based $izY$ is largely based on the slightly redder cutoff of its $i$-band blue edge relative to comparable instruments ($\sim$0.71$~\mu$m on DECam as opposed to $\sim$0.69~$\mu$m on SDSS, Pan-STARRS, and Rubin).
As the EIRTF library's blue cutoff is 0.7~$\mu$m, the DECam $i$-band allows us to compute higher fidelity synthetic photometry for these spectra.
Transformations between DECam and other facilities may be found in \citet[][Appendix B]{2021ApJS..255...20A}.

On a similar note, we could equally have chosen the F814W filter on WFC3/UVIS as opposed to ACS/WFC, but the ACS/WFC one is more widely used, especially for the distance-scale applications that motivated this work. Differences between these respective filters are generally within $\pm$0.02~mag, as detailed in \citet{2018wfc..rept....2D}.

\added{We caution that there are a number of instrumental effects which may have subtle impacts on measured stellar colors and magnitudes, thereby introducing systematics not accounted for in this analysis. 
One such effect that is particularly relevant for \Roman and \Euclid is angle of incidence (AOI) variations, where the angle at which incoming light enters the filter and/or detector medium causes variations in the effective passband between the center and edges of the field of view \citep{2020JATIS...6d6001M, 2022A&A...662A..92E}. \citet{2022A&A...662A..92E} conduct a detailed examination of passband variations across the \Euclid FOV, and determine that the net impact on photometry of individual sources is of the order 1-5 millimag. 
While similar quantitative studies are not yet published for \Roman, we anticipate roughly comparable results for WFI. 
We also note that \JWST/NIRCam shows substantial chip-to-chip variations in overall throughput \citep{2022RNAAS...6..191B} as well as subtler differences in filter curve shapes\footnote{\url{https://jwst-docs.stsci.edu/jwst-near-infrared-camera/nircam-instrumentation/nircam-filters}} due to quantum efficiency variations. 
A detailed study of the impacts of spatial transmission variations on color transformations and residuals is beyond the scope of this paper, but we urge readers to keep such effects in mind, especially when high-precision predictions are needed.}

Finally, we note that \JWST calibration continues to evolve as more in-flight data becomes available \citep[e.g.][]{2022RNAAS...6..191B, 2023PASP..135d8002R, 2024PASP..136b4501M}, and the same is expected for \Euclid and \Roman as the missions mature.

\subsection{Synthetic photometry} \label{ssec:method}

For a spectrum of flux density $\mathbf{f_{\lambda}}$ integrated over a set of bandpasses $\mathbf{P}$ sampled on the same wavelength grid, 
the integrated flux densities $\mathbf{f}_P$ 
are simply:
\begin{equation} \label{eqn:flux_dot}
    \mathbf{f}_P = \mathbf{P} \cdot \mathbf{f}_\lambda
\end{equation}
and the covariance between them is:
\begin{equation} \label{eqn:flux_cov}
    \Cov[{\mathbf{f}_P}] = \mathbf{P} \cdot \Cov[{\mathbf{f}_\lambda}] \cdot \mathbf{P}^T
\end{equation}
as in \citet{2022arXiv220606215G} and \citet{2023AJ....166..236D}.
Individual output magnitudes and uncertainties are available upon request.

We use the CALSPEC file \texttt{alpha\_lyr\_stis\_011.fits} to calculate our fiducial Vega zeropoints for all bands.
We also provide conversions to the ABmag and STmag systems in \autoref{tab:zp_conv} for convenience, and to the ``Sirius-Vega"\added{\footnote{Note that this name is hyphenated, whereas in \autoref{tab:zp_conv} the ``Sirius--Vega" column denotes the difference between the ``Sirius-Vega" and ``Vega-Vega" magnitude systems with the requisite en dash.}} system used for \JWST \citep{2022AJ....163...45R} as well, \added{which retains Vega as the overall magnitude standard but adopts Sirius as the infrared color standard due to Vega's IR excess}.
Note that for NIRCam, AB magnitudes are redefined to depend solely on pixel area, such that all NIRCam short-wavelength channel filters have nearly the same AB zeropoint; thus, their AB-Vega conversions differ substantially from those for other instruments.
All other AB zeropoints are calculated using the traditional definition \citep{1974ApJS...27...21O}.

\begin{table}
    \centering
    \begin{tabular}{lrrrr}
    \hline \hline
Band & Pivot & AB--Vega & ST--Vega & Sirius--Vega \\
     & [$\mu$m] & [mag] & [mag] & [mag] \\
\hline
\multicolumn{5}{c}{\Euclid/NISP} \\
$Y_E$ & 1.079 & 0.694 & 2.166 & 0.027 \\
$J_E$ & 1.362 & 1.058 & 3.037 & 0.023 \\
$H_E$ & 1.765 & 1.493 & 4.034 & 0.014 \\
\hline
\multicolumn{5}{c}{CTIO/DECam, OGLE IV, 2MASS} \\
$i$ & 0.781 & 0.418 & 1.190 & 0.055 \\
$I$ & 0.790 & 0.441 & 1.237 & 0.054 \\
$z$ & 0.917 & 0.527 & 1.647 & 0.030 \\
$Y$ & 0.990 & 0.582 & 1.867 & 0.025 \\
$J$ & 1.239 & 0.913 & 2.687 & 0.026 \\
$H$ & 1.649 & 1.391 & 3.786 & 0.015 \\
$K_S$ & 2.164 & 1.864 & 4.848 & 0.011 \\
\hline
\multicolumn{5}{c}{\HST ACS/WFC + WFC3/IR} \\
F814W & 0.805 & 0.431 & 1.267 & 0.050 \\
F110W & 1.153 & 0.777 & 2.394 & 0.026 \\
F125W & 1.249 & 0.920 & 2.710 & 0.026 \\
F160W & 1.537 & 1.274 & 3.515 & 0.019 \\
\hline
\multicolumn{5}{c}{\JWST/NIRCam} \\
F090W & 0.903 & 1.821 & 1.608 & 0.034 \\
F115W & 1.154 & 2.015 & 2.413 & 0.027 \\
F150W & 1.501 & 2.126 & 3.419 & 0.020 \\
F200W & 1.988 & 2.422 & 4.500 & 0.011 \\
\hline
\multicolumn{5}{c}{\Roman/WFI} \\
F087 & 0.870 & 0.497 & 1.501 & 0.040 \\
F106 & 1.057 & 0.664 & 2.091 & 0.027 \\
F129 & 1.290 & 0.972 & 2.833 & 0.025 \\
F158 & 1.575 & 1.305 & 3.599 & 0.018 \\
F184 & 1.839 & 1.571 & 4.203 & 0.012 \\
F213 & 2.123 & 1.826 & 4.768 & 0.011 \\
\hline
    \end{tabular}
    \caption{Pivot wavelengths and conversions from Vega magnitudes to the AB, ST, and Sirius-Vega systems for all filters. These conversion factors should be added to the Vega magnitudes reported here to obtain transformations and photometry in the desired system.
    }
    \label{tab:zp_conv}
\end{table}

\section{Results} \label{sec:res}

\subsection{Transformation coefficients}
We fit continuous piecewise linear transformations using the Python library \texttt{pwlf} \citep{pwlf}, which take the form
\begin{align}
Y = \begin{cases}
      c_0 + c_1(X-X_0) & X_0 \leq X \leq X_1 \\
      c_0 + c_1(X-X_0) + c_2(X-X_1) & X_1 < X \leq X_2 \\
      \vdots & \vdots \\
      \sum\limits_{i=1}^{n} c_{i-1} + c_{i}(X-X_{i-1}) & X_{n-1} < X \leq X_{n}
\end{cases} \label{eq:pwlf}
\end{align}
for a filter pair difference $Y$, color term $X$, and number of line segments $n-1$.

In \citet{2023AJ....166..236D} we chose the number of line segments for each filter combination visually.
In this work, we find the best-fit number of line segments with Bayesian optimization using the \texttt{GPyOpt} library \citep{gpyopt2016}, following the example in the \texttt{pwlf} documentation.\footnote{\url{https://jekel.me/piecewise\_linear\_fit\_py/examples.html\#find-the-best-number-of-line-segments}}
We set the regularization parameter $l$ to the minimum uncertainty on $Y$, and impose the additional constraints that all segments must span at least 0.1 mag in color and have at least 25 stars in their color range.
In addition to the global rejection criteria described in \citet{2023AJ....166..236D} Section 4.1 (items 2-5), we require stars to have total uncertainties (random and systematic \added{combined in quadrature}) less than 0.05 mag in the three unique filters used in a given transformation.
\added{The global rejection criteria yield a fiducial sample of 856 spectra of 716 stars, or about 2/3 of the total library sample, and the additional uncertainty criteria yield between 605 and 815 spectra per filter combination as transformation inputs.}

We list all transformation coefficients in Tables~\ref{tab:fits_euclid} through \ref{tab:fits_roman}, and show figures of all fits in \autoref{app:fit_figs}.
\added{We also provide the residual RMS scatter for stars within the $X$-range of a given line segment, $\sigma_{\mathrm{seg}}$, the number of stars per segment $N_{\star}$, and the overall residual scatter of all stars across all segments, $\sigma_{\mathrm{fit}}$.}

\begin{longtable*}{llRRRRRRR}
    \caption{Transformations to the \Euclid/NISP filter system.}
    \label{tab:fits_euclid} \\
    \hline \hline
\multicolumn{1}{c}{$Y$} & \multicolumn{1}{c}{$X$} & X_0 & X_1 & \multicolumn{1}{C}{c_0} & \multicolumn{1}{C}{c_1} & \multicolumn{1}{C}{\sigma_\mathrm{seg}} & N_\star & \multicolumn{1}{C}{\sigma_\mathrm{fit}} \\ \hline
\endfirsthead
\hline
\multicolumn{1}{c}{$Y$} & \multicolumn{1}{c}{$X$} & X_0 & X_1 & \multicolumn{1}{C}{c_0} & \multicolumn{1}{C}{c_1} & \multicolumn{1}{C}{\sigma_\mathrm{seg}} & N_\star & \multicolumn{1}{C}{\sigma_\mathrm{fit}} \\ \hline
\endhead
\hline
\multicolumn{9}{c}{Continued on next page} \\
\endfoot
\hline
\endlastfoot
\multicolumn{9}{c}{From CTIO/DECam + 2MASS} \\
$Y_E$--$Y$ & $i$--$Y$ & -0.35 & -0.02 & 0.000 & -0.125 & 0.013 & 34 & 0.013 \\
 &  & -0.02 & 0.14 & -0.007 & -0.585 & 0.015 & 153 & 0.013 \\
 &  & 0.14 & 1.08 & -0.055 & -0.253 & 0.012 & 418 & 0.013 \\
$Y_E$--$J$ & $J$--$H$ & -0.13 & 0.21 & -0.001 & 0.663 & 0.012 & 163 & 0.013 \\
 &  & 0.21 & 1.03 & 0.051 & 0.422 & 0.013 & 598 & 0.013 \\
 & $J$--$K_S$ & -0.29 & -0.03 & -0.008 & 0.287 & 0.010 & 30 & 0.013 \\
 &  & -0.03 & 0.25 & 0.001 & 0.596 & 0.016 & 142 & 0.013 \\
 &  & 0.25 & 1.35 & 0.067 & 0.330 & 0.012 & 589 & 0.013 \\
$J_E$--$J$ & $J$--$H$ & -0.13 & 0.35 & -0.007 & -0.340 & 0.007 & 339 & 0.007 \\
 &  & 0.35 & 1.03 & -0.029 & -0.279 & 0.007 & 441 & 0.007 \\
 & $J$--$K_S$ & -0.29 & -0.00 & -0.007 & -0.166 & 0.007 & 38 & 0.009 \\
 &  & -0.00 & 0.28 & -0.007 & -0.311 & 0.010 & 164 & 0.009 \\
 &  & 0.28 & 0.71 & -0.026 & -0.246 & 0.009 & 435 & 0.009 \\
 &  & 0.71 & 1.35 & -0.080 & -0.170 & 0.011 & 143 & 0.009 \\
$H_E$--$H$ & $J$--$H$ & -0.13 & -0.03 & -0.006 & -0.266 & 0.009 & 28 & 0.013 \\
 &  & -0.03 & 0.42 & -0.001 & -0.045 & 0.013 & 368 & 0.013 \\
 &  & 0.42 & 1.03 & 0.026 & -0.107 & 0.013 & 382 & 0.013 \\

\hline \multicolumn{9}{c}{From \HST WFC3/IR} \\
$Y_E$--F110W & F110W--F160W & -0.17 & 0.09 & 0.003 & 0.117 & 0.003 & 70 & 0.006 \\
 &  & 0.09 & 1.09 & -0.003 & 0.187 & 0.006 & 691 & 0.006 \\
$Y_E$--F125W & F125W--F160W & -0.09 & 0.15 & -0.003 & 0.858 & 0.011 & 138 & 0.015 \\
 &  & 0.15 & 0.54 & 0.037 & 0.596 & 0.015 & 571 & 0.015 \\
 &  & 0.54 & 0.76 & -0.053 & 0.762 & 0.025 & 52 & 0.015 \\
$J_E$--F110W & F110W--F160W & -0.17 & 0.26 & 0.001 & -0.635 & 0.006 & 160 & 0.006 \\
 &  & 0.26 & 1.09 & -0.022 & -0.547 & 0.007 & 613 & 0.006 \\
$J_E$--F125W & F125W--F160W & -0.09 & 0.76 & -0.004 & -0.396 & 0.004 & 778 & 0.004 \\
$J_E$--F160W & F814W--F160W & -0.56 & 0.06 & 0.004 & 0.106 & 0.009 & 48 & 0.011 \\
 &  & 0.06 & 1.43 & -0.003 & 0.219 & 0.010 & 510 & 0.011 \\
 &  & 1.43 & 2.29 & 0.105 & 0.143 & 0.021 & 48 & 0.011 \\
 & F110W--F160W & -0.17 & 0.26 & 0.001 & 0.364 & 0.005 & 159 & 0.006 \\
 &  & 0.26 & 1.09 & -0.022 & 0.453 & 0.007 & 614 & 0.006 \\
 & F125W--F160W & -0.09 & 0.76 & -0.004 & 0.604 & 0.004 & 778 & 0.004 \\
$H_E$--F160W & F814W--F160W & -0.56 & 0.69 & 0.006 & -0.099 & 0.012 & 233 & 0.015 \\
 &  & 0.69 & 2.29 & 0.058 & -0.174 & 0.016 & 372 & 0.015 \\
 & F110W--F160W & -0.17 & -0.00 & -0.002 & -0.426 & 0.011 & 37 & 0.015 \\
 &  & -0.00 & 0.32 & -0.001 & -0.135 & 0.009 & 165 & 0.015 \\
 &  & 0.32 & 0.68 & 0.067 & -0.351 & 0.016 & 446 & 0.015 \\
 &  & 0.68 & 1.09 & 0.160 & -0.487 & 0.021 & 124 & 0.015 \\
 & F125W--F160W & -0.09 & 0.26 & 0.016 & -0.316 & 0.013 & 283 & 0.018 \\
 &  & 0.26 & 0.50 & 0.058 & -0.480 & 0.019 & 391 & 0.018 \\
 &  & 0.50 & 0.76 & 0.162 & -0.687 & 0.025 & 102 & 0.018 \\

\hline \multicolumn{9}{c}{From \JWST/NIRCam} \\
$Y_E$--F115W & F115W--F150W & -0.13 & 0.19 & -0.006 & 0.299 & 0.006 & 127 & 0.007 \\
 &  & 0.19 & 0.68 & 0.008 & 0.229 & 0.006 & 588 & 0.007 \\
 &  & 0.68 & 0.96 & -0.072 & 0.346 & 0.011 & 46 & 0.007 \\
 & F115W--F200W & -0.27 & 0.33 & 0.003 & 0.196 & 0.007 & 180 & 0.007 \\
 &  & 0.33 & 1.53 & 0.021 & 0.141 & 0.006 & 579 & 0.007 \\
$J_E$--F115W & F115W--F150W & -0.13 & 0.27 & -0.000 & -0.644 & 0.003 & 186 & 0.004 \\
 &  & 0.27 & 0.96 & -0.013 & -0.595 & 0.005 & 588 & 0.004 \\
 & F115W--F200W & -0.27 & 0.01 & -0.003 & -0.307 & 0.008 & 39 & 0.009 \\
 &  & 0.01 & 0.32 & -0.001 & -0.513 & 0.010 & 130 & 0.009 \\
 &  & 0.32 & 0.82 & -0.046 & -0.370 & 0.009 & 432 & 0.009 \\
 &  & 0.82 & 1.53 & -0.102 & -0.301 & 0.010 & 170 & 0.009 \\
$J_E$--F150W & F090W--F150W & -0.38 & 0.15 & 0.009 & 0.146 & 0.006 & 75 & 0.008 \\
 &  & 0.15 & 1.64 & -0.003 & 0.220 & 0.009 & 551 & 0.008 \\
 & F115W--F150W & -0.13 & 0.27 & -0.000 & 0.356 & 0.003 & 186 & 0.004 \\
 &  & 0.27 & 0.96 & -0.014 & 0.406 & 0.005 & 588 & 0.004 \\
 & F150W--F200W & -0.17 & -0.03 & 0.004 & 0.343 & 0.008 & 28 & 0.018 \\
 &  & -0.03 & 0.08 & 0.022 & 0.959 & 0.018 & 183 & 0.018 \\
 &  & 0.08 & 0.23 & 0.046 & 0.673 & 0.016 & 336 & 0.018 \\
 &  & 0.23 & 0.57 & 0.101 & 0.429 & 0.021 & 253 & 0.018 \\
$H_E$--F150W & F090W--F150W & -0.38 & 0.49 & 0.009 & -0.164 & 0.010 & 176 & 0.015 \\
 &  & 0.49 & 1.64 & 0.068 & -0.283 & 0.016 & 449 & 0.015 \\
 & F115W--F150W & -0.13 & 0.01 & 0.008 & -0.561 & 0.012 & 37 & 0.015 \\
 &  & 0.01 & 0.26 & 0.003 & -0.237 & 0.010 & 149 & 0.015 \\
 &  & 0.26 & 0.60 & 0.071 & -0.493 & 0.015 & 457 & 0.015 \\
 &  & 0.60 & 0.96 & 0.178 & -0.672 & 0.019 & 129 & 0.015 \\
 & F150W--F200W & -0.17 & 0.57 & -0.011 & -0.744 & 0.010 & 815 & 0.010 \\
$H_E$--F200W & F090W--F200W & -0.51 & -0.12 & 0.006 & 0.134 & 0.010 & 30 & 0.012 \\
 &  & -0.12 & 1.02 & -0.005 & 0.039 & 0.012 & 352 & 0.012 \\
 &  & 1.02 & 2.14 & -0.053 & 0.085 & 0.012 & 242 & 0.012 \\
 & F115W--F200W & -0.27 & -0.05 & 0.001 & 0.242 & 0.008 & 27 & 0.011 \\
 &  & -0.05 & 0.77 & -0.009 & 0.067 & 0.011 & 533 & 0.011 \\
 &  & 0.77 & 1.53 & -0.072 & 0.149 & 0.013 & 211 & 0.011 \\
 & F150W--F200W & -0.17 & -0.03 & 0.004 & 0.484 & 0.006 & 27 & 0.009 \\
 &  & -0.03 & 0.25 & -0.004 & 0.209 & 0.009 & 574 & 0.009 \\
 &  & 0.25 & 0.57 & -0.029 & 0.311 & 0.009 & 214 & 0.009 \\ \hline
\end{longtable*}

\begin{longtable*}{llRRRRRRR}
    \caption{Transformations to the \JWST/NIRCam filter system.}
    \label{tab:roman_jwst_fits} \\
    \hline \hline
\multicolumn{1}{c}{$Y$} & \multicolumn{1}{c}{$X$} & X_0 & X_1 & \multicolumn{1}{C}{c_0} & \multicolumn{1}{C}{c_1} & \multicolumn{1}{C}{\sigma_\mathrm{seg}} & N_\star & \multicolumn{1}{C}{\sigma_\mathrm{fit}} \\ \hline
\endfirsthead
\hline
\multicolumn{1}{c}{$Y$} & \multicolumn{1}{c}{$X$} & X_0 & X_1 & \multicolumn{1}{C}{c_0} & \multicolumn{1}{C}{c_1} & \multicolumn{1}{C}{\sigma_\mathrm{seg}} & N_\star & \multicolumn{1}{C}{\sigma_\mathrm{fit}} \\ \hline
\endhead
\hline
\multicolumn{9}{c}{Continued on next page} \\
\endfoot
\hline
\endlastfoot
\multicolumn{9}{c}{From \Euclid/NISP} \\
F115W--$Y_E$ & $Y_E$--$H_E$ & -0.25 & 0.34 & 0.001 & -0.181 & 0.005 & 156 & 0.006 \\
 &  & 0.34 & 1.63 & -0.014 & -0.136 & 0.006 & 604 & 0.006 \\
 & $Y_E$--$J_E$ & -0.13 & 0.85 & 0.003 & -0.292 & 0.005 & 761 & 0.005 \\
F115W--$J_E$ & $J_E$--$H_E$ & -0.12 & 0.01 & 0.004 & 0.655 & 0.010 & 37 & 0.013 \\
 &  & 0.01 & 0.15 & 0.002 & 1.119 & 0.015 & 137 & 0.013 \\
 &  & 0.15 & 0.41 & 0.067 & 0.671 & 0.012 & 412 & 0.013 \\
 &  & 0.41 & 0.78 & 0.116 & 0.551 & 0.012 & 186 & 0.013 \\
 & $Y_E$--$J_E$ & -0.13 & 0.85 & 0.003 & 0.708 & 0.005 & 761 & 0.005 \\
F150W--$J_E$ & $J_E$--$H_E$ & -0.12 & 0.26 & -0.011 & -0.516 & 0.006 & 368 & 0.008 \\
 &  & 0.26 & 0.78 & -0.039 & -0.411 & 0.009 & 433 & 0.008 \\
 & $Y_E$--$J_E$ & -0.13 & 0.21 & -0.002 & -0.380 & 0.006 & 160 & 0.008 \\
 &  & 0.21 & 0.85 & 0.020 & -0.481 & 0.009 & 601 & 0.008 \\
F150W--$H_E$ & $J_E$--$H_E$ & -0.12 & 0.26 & -0.011 & 0.484 & 0.006 & 368 & 0.008 \\
 &  & 0.26 & 0.78 & -0.039 & 0.590 & 0.009 & 433 & 0.008 \\
 & $Y_E$--$H_E$ & -0.25 & 0.55 & -0.016 & 0.209 & 0.009 & 317 & 0.011 \\
 &  & 0.55 & 1.63 & -0.074 & 0.315 & 0.012 & 443 & 0.011 \\
F200W--$H_E$ & $J_E$--$H_E$ & -0.12 & 0.47 & 0.013 & -0.153 & 0.013 & 686 & 0.013 \\
 &  & 0.47 & 0.78 & 0.083 & -0.301 & 0.016 & 114 & 0.013 \\
 & $Y_E$--$H_E$ & -0.25 & -0.05 & -0.003 & -0.269 & 0.011 & 29 & 0.012 \\
 &  & -0.05 & 0.82 & 0.007 & -0.056 & 0.011 & 485 & 0.012 \\
 &  & 0.82 & 1.63 & 0.078 & -0.143 & 0.013 & 245 & 0.012 \\

\hline \multicolumn{9}{c}{From CTIO/DECam, OGLE IV, 2MASS} \\
F090W--$i$ & $i$--$Y$ & -0.35 & -0.03 & 0.001 & -0.551 & 0.004 & 34 & 0.007 \\
 &  & -0.03 & 1.08 & -0.001 & -0.643 & 0.007 & 571 & 0.007 \\
 & $i$--$z$ & -0.23 & 0.78 & -0.008 & -0.892 & 0.003 & 605 & 0.003 \\
F090W--$I$ & $I$--$K_S$ & -0.80 & -0.02 & -0.001 & -0.320 & 0.017 & 37 & 0.015 \\
 &  & -0.02 & 0.62 & 0.002 & -0.123 & 0.011 & 126 & 0.015 \\
 &  & 0.62 & 1.51 & 0.024 & -0.158 & 0.011 & 353 & 0.015 \\
 &  & 1.51 & 2.79 & 0.113 & -0.217 & 0.026 & 96 & 0.015 \\
F090W--$z$ & $i$--$z$ & -0.23 & 0.78 & -0.008 & 0.108 & 0.003 & 605 & 0.003 \\
F090W--$Y$ & $i$--$Y$ & -0.35 & -0.04 & 0.003 & 0.453 & 0.004 & 33 & 0.007 \\
 &  & -0.04 & 1.08 & -0.001 & 0.357 & 0.007 & 572 & 0.007 \\
F115W--$J$ & $J$--$H$ & -0.13 & 0.22 & 0.002 & 0.348 & 0.006 & 171 & 0.006 \\
 &  & 0.22 & 1.03 & 0.030 & 0.219 & 0.006 & 603 & 0.006 \\
 & $J$--$K_S$ & -0.29 & 0.42 & 0.011 & 0.247 & 0.008 & 349 & 0.007 \\
 &  & 0.42 & 1.35 & 0.049 & 0.158 & 0.006 & 425 & 0.007 \\
F150W--$H$ & $J$--$H$ & -0.13 & 0.35 & -0.011 & 0.283 & 0.009 & 337 & 0.010 \\
 &  & 0.35 & 1.03 & -0.045 & 0.379 & 0.010 & 443 & 0.010 \\
F200W--$K_S$ & $I$--$K_S$ & -0.80 & 2.79 & -0.010 & 0.017 & 0.017 & 610 & 0.017 \\
 & $J$--$K_S$ & -0.29 & 0.23 & -0.012 & 0.085 & 0.013 & 160 & 0.017 \\
 &  & 0.23 & 0.70 & 0.005 & 0.009 & 0.017 & 471 & 0.017 \\
 &  & 0.70 & 1.35 & -0.045 & 0.080 & 0.018 & 146 & 0.017 \\

\hline \multicolumn{9}{c}{From \HST ACS/WFC + WFC3/IR} \\
F090W--F814W & F814W--F160W & -0.56 & 0.01 & 0.007 & -0.293 & 0.012 & 41 & 0.012 \\
 &  & 0.01 & 0.57 & 0.005 & -0.142 & 0.009 & 126 & 0.012 \\
 &  & 0.57 & 1.29 & 0.027 & -0.179 & 0.009 & 351 & 0.012 \\
 &  & 1.29 & 2.29 & 0.173 & -0.293 & 0.024 & 88 & 0.012 \\
F115W--F110W & F110W--F160W & -0.17 & 0.17 & 0.008 & -0.114 & 0.005 & 107 & 0.005 \\
 &  & 0.17 & 1.09 & -0.007 & -0.023 & 0.005 & 665 & 0.005 \\
F115W--F125W & F125W--F160W & -0.09 & 0.16 & 0.001 & 0.455 & 0.006 & 142 & 0.009 \\
 &  & 0.16 & 0.76 & 0.021 & 0.326 & 0.009 & 632 & 0.009 \\
F150W--F160W & F814W--F160W & -0.56 & -0.03 & -0.005 & 0.007 & 0.004 & 35 & 0.005 \\
 &  & -0.03 & 2.29 & -0.004 & 0.045 & 0.005 & 571 & 0.005 \\
 & F110W--F160W & -0.17 & 1.09 & -0.004 & 0.087 & 0.005 & 773 & 0.005 \\
 & F125W--F160W & -0.09 & 0.76 & -0.004 & 0.124 & 0.004 & 778 & 0.004 \\ \hline
\end{longtable*}

\begin{longtable*}{llRRRRRRR}
    \caption{Transformations to the \Roman/WFI filter system.}
    \label{tab:fits_roman} \\
    \hline \hline
\multicolumn{1}{c}{$Y$} & \multicolumn{1}{c}{$X$} & X_0 & X_1 & \multicolumn{1}{C}{c_0} & \multicolumn{1}{C}{c_1} & \multicolumn{1}{C}{\sigma_\mathrm{seg}} & N_\star & \multicolumn{1}{C}{\sigma_\mathrm{fit}} \\ \hline
\endfirsthead
\hline
\multicolumn{1}{c}{$Y$} & \multicolumn{1}{c}{$X$} & X_0 & X_1 & \multicolumn{1}{C}{c_0} & \multicolumn{1}{C}{c_1} & \multicolumn{1}{C}{\sigma_\mathrm{seg}} & N_\star & \multicolumn{1}{C}{\sigma_\mathrm{fit}} \\ \hline
\endhead
\hline
\multicolumn{9}{c}{Continued on next page} \\
\endfoot
\hline
\endlastfoot
\multicolumn{9}{c}{From \Euclid/NISP} \\
F106--$Y_E$ & $Y_E$--$H_E$ & -0.25 & 1.63 & 0.000 & 0.046 & 0.003 & 759 & 0.003 \\
 & $Y_E$--$J_E$ & -0.13 & 0.85 & -0.002 & 0.092 & 0.003 & 760 & 0.003 \\
F129--$J_E$ & $J_E$--$H_E$ & -0.12 & 0.26 & 0.007 & 0.275 & 0.004 & 367 & 0.005 \\
 &  & 0.26 & 0.78 & 0.025 & 0.205 & 0.005 & 415 & 0.005 \\
 & $Y_E$--$J_E$ & -0.13 & 0.16 & 0.002 & 0.197 & 0.003 & 119 & 0.004 \\
 &  & 0.16 & 0.85 & -0.006 & 0.244 & 0.005 & 642 & 0.004 \\
F158--$J_E$ & $J_E$--$H_E$ & -0.12 & 0.26 & -0.007 & -0.694 & 0.007 & 368 & 0.009 \\
 &  & 0.26 & 0.78 & -0.025 & -0.624 & 0.010 & 433 & 0.009 \\
 & $Y_E$--$J_E$ & -0.13 & 0.24 & 0.004 & -0.502 & 0.009 & 182 & 0.012 \\
 &  & 0.24 & 0.85 & 0.054 & -0.712 & 0.013 & 579 & 0.012 \\
F158--$H_E$ & $J_E$--$H_E$ & -0.12 & 0.26 & -0.007 & 0.308 & 0.007 & 367 & 0.009 \\
 &  & 0.26 & 0.78 & -0.025 & 0.377 & 0.010 & 434 & 0.009 \\
 & $Y_E$--$H_E$ & -0.25 & 0.55 & -0.010 & 0.133 & 0.009 & 320 & 0.010 \\
 &  & 0.55 & 1.63 & -0.048 & 0.201 & 0.011 & 440 & 0.010 \\
F184--$H_E$ & $J_E$--$H_E$ & -0.12 & -0.02 & -0.008 & -0.323 & 0.006 & 29 & 0.009 \\
 &  & -0.02 & 0.18 & -0.003 & -0.026 & 0.007 & 185 & 0.009 \\
 &  & 0.18 & 0.78 & 0.017 & -0.133 & 0.010 & 585 & 0.009 \\
 & $Y_E$--$H_E$ & -0.25 & -0.02 & -0.006 & -0.156 & 0.007 & 30 & 0.009 \\
 &  & -0.02 & 0.46 & -0.003 & -0.011 & 0.007 & 197 & 0.009 \\
 &  & 0.46 & 1.63 & 0.024 & -0.071 & 0.010 & 531 & 0.009 \\

\hline \multicolumn{9}{c}{From CTIO/DECam, OGLE IV, 2MASS} \\
F087--$i$ & $i$--$Y$ & -0.35 & -0.02 & 0.001 & -0.390 & 0.004 & 34 & 0.006 \\
 &  & -0.02 & 1.08 & -0.002 & -0.489 & 0.006 & 571 & 0.006 \\
 & $i$--$z$ & -0.23 & -0.04 & -0.002 & -0.602 & 0.003 & 30 & 0.003 \\
 &  & -0.04 & 0.78 & -0.006 & -0.685 & 0.003 & 575 & 0.003 \\
F087--$I$ & $I$--$K_S$ & -0.80 & -0.10 & -0.009 & -0.257 & 0.015 & 29 & 0.011 \\
 &  & -0.10 & 1.36 & 0.007 & -0.106 & 0.008 & 429 & 0.011 \\
 &  & 1.36 & 2.79 & 0.064 & -0.147 & 0.017 & 154 & 0.011 \\
F087--$z$ & $i$--$z$ & -0.23 & -0.05 & -0.002 & 0.399 & 0.003 & 30 & 0.003 \\
 &  & -0.05 & 0.78 & -0.006 & 0.314 & 0.003 & 575 & 0.003 \\
F106--$z$ & $i$--$z$ & -0.23 & -0.03 & -0.003 & -0.646 & 0.017 & 33 & 0.017 \\
 &  & -0.03 & 0.08 & -0.016 & -1.065 & 0.022 & 122 & 0.017 \\
 &  & 0.08 & 0.30 & -0.045 & -0.687 & 0.016 & 392 & 0.017 \\
 &  & 0.30 & 0.78 & -0.110 & -0.474 & 0.016 & 58 & 0.017 \\
F106--$Y$ & $i$--$Y$ & -0.35 & -0.02 & 0.002 & -0.080 & 0.010 & 34 & 0.011 \\
 &  & -0.02 & 0.13 & -0.004 & -0.448 & 0.012 & 140 & 0.011 \\
 &  & 0.13 & 1.08 & -0.039 & -0.184 & 0.010 & 431 & 0.011 \\
F129--$J$ & $J$--$H$ & -0.13 & 0.36 & -0.004 & -0.139 & 0.007 & 340 & 0.007 \\
 &  & 0.36 & 1.03 & -0.015 & -0.107 & 0.008 & 440 & 0.007 \\
 & $J$--$K_S$ & -0.29 & 0.67 & -0.009 & -0.105 & 0.007 & 609 & 0.007 \\
 &  & 0.67 & 1.35 & -0.038 & -0.062 & 0.009 & 171 & 0.007 \\
F158--$H$ & $J$--$H$ & -0.13 & 0.33 & -0.005 & 0.155 & 0.006 & 307 & 0.006 \\
 &  & 0.33 & 1.03 & -0.021 & 0.205 & 0.006 & 473 & 0.006 \\
F184--$H$ & $J$--$H$ & -0.13 & -0.02 & -0.011 & -0.535 & 0.018 & 28 & 0.023 \\
 &  & -0.02 & 0.39 & -0.000 & -0.086 & 0.023 & 336 & 0.023 \\
 &  & 0.39 & 1.03 & 0.051 & -0.218 & 0.024 & 412 & 0.023 \\
F213--$K_S$ & $I$--$K_S$ & -0.80 & 2.79 & -0.001 & 0.005 & 0.004 & 612 & 0.004 \\
 & $J$--$K_S$ & -0.29 & 1.35 & -0.001 & 0.010 & 0.004 & 780 & 0.004 \\

\hline \multicolumn{9}{c}{From \HST ACS/WFC + WFC3/IR} \\
F087--F814W & F814W--F160W & -0.56 & -0.03 & 0.006 & -0.185 & 0.008 & 35 & 0.009 \\
 &  & -0.03 & 1.25 & 0.008 & -0.118 & 0.006 & 470 & 0.009 \\
 &  & 1.25 & 2.29 & 0.114 & -0.203 & 0.016 & 101 & 0.009 \\
F106--F110W & F110W--F160W & -0.17 & 0.74 & 0.003 & 0.242 & 0.007 & 699 & 0.007 \\
 &  & 0.74 & 1.09 & -0.070 & 0.340 & 0.011 & 61 & 0.007 \\
F106--F125W & F125W--F160W & -0.09 & 0.14 & -0.005 & 1.000 & 0.013 & 131 & 0.018 \\
 &  & 0.14 & 0.54 & 0.039 & 0.683 & 0.017 & 576 & 0.018 \\
 &  & 0.54 & 0.76 & -0.085 & 0.912 & 0.031 & 53 & 0.018 \\
F129--F110W & F110W--F160W & -0.17 & 0.22 & 0.003 & -0.487 & 0.006 & 135 & 0.008 \\
 &  & 0.22 & 1.09 & -0.024 & -0.367 & 0.008 & 638 & 0.008 \\
F129--F125W & F125W--F160W & -0.09 & 0.76 & -0.003 & -0.146 & 0.003 & 778 & 0.003 \\
F158--F160W & F814W--F160W & -0.56 & 0.77 & 0.003 & -0.013 & 0.002 & 277 & 0.003 \\
 &  & 0.77 & 2.29 & 0.022 & -0.037 & 0.004 & 329 & 0.003 \\
 & F110W--F160W & -0.17 & 0.46 & 0.004 & -0.029 & 0.003 & 375 & 0.004 \\
 &  & 0.46 & 1.09 & 0.033 & -0.092 & 0.006 & 398 & 0.004 \\
 & F125W--F160W & -0.09 & 0.32 & 0.004 & -0.044 & 0.003 & 379 & 0.005 \\
 &  & 0.32 & 0.76 & 0.029 & -0.123 & 0.006 & 399 & 0.005 \\

\hline \multicolumn{9}{c}{From \JWST/NIRCam} \\
F087--F090W & F090W--F150W & -0.38 & -0.00 & -0.006 & 0.141 & 0.006 & 38 & 0.005 \\
 &  & -0.00 & 0.91 & -0.007 & 0.060 & 0.004 & 413 & 0.005 \\
 &  & 0.91 & 1.64 & -0.049 & 0.106 & 0.007 & 169 & 0.005 \\
 & F090W--F200W & -0.51 & -0.01 & -0.005 & 0.103 & 0.006 & 38 & 0.005 \\
 &  & -0.01 & 1.28 & -0.006 & 0.048 & 0.004 & 472 & 0.005 \\
 &  & 1.28 & 2.14 & -0.045 & 0.078 & 0.009 & 108 & 0.005 \\
F106--F090W & F090W--F150W & -0.38 & 0.03 & 0.016 & -0.480 & 0.014 & 45 & 0.012 \\
 &  & 0.03 & 0.75 & 0.008 & -0.236 & 0.011 & 296 & 0.012 \\
 &  & 0.75 & 1.64 & 0.072 & -0.321 & 0.013 & 285 & 0.012 \\
 & F090W--F200W & -0.51 & 0.06 & 0.014 & -0.355 & 0.016 & 48 & 0.014 \\
 &  & 0.06 & 0.90 & 0.003 & -0.187 & 0.012 & 284 & 0.014 \\
 &  & 0.90 & 2.14 & 0.035 & -0.223 & 0.015 & 292 & 0.014 \\
F106--F115W & F115W--F150W & -0.13 & 0.18 & -0.008 & 0.397 & 0.009 & 123 & 0.009 \\
 &  & 0.18 & 0.68 & 0.010 & 0.299 & 0.009 & 589 & 0.009 \\
 &  & 0.68 & 0.96 & -0.111 & 0.476 & 0.016 & 48 & 0.009 \\
 & F115W--F200W & -0.27 & 0.32 & 0.003 & 0.257 & 0.010 & 170 & 0.009 \\
 &  & 0.32 & 1.53 & 0.025 & 0.187 & 0.009 & 588 & 0.009 \\
F129--F115W & F115W--F150W & -0.13 & 0.25 & 0.000 & -0.454 & 0.004 & 175 & 0.007 \\
 &  & 0.25 & 0.96 & -0.016 & -0.389 & 0.007 & 599 & 0.007 \\
 & F115W--F200W & -0.27 & 0.01 & -0.002 & -0.223 & 0.006 & 39 & 0.008 \\
 &  & 0.01 & 0.32 & -0.001 & -0.357 & 0.008 & 129 & 0.008 \\
 &  & 0.32 & 0.83 & -0.037 & -0.242 & 0.007 & 442 & 0.008 \\
 &  & 0.83 & 1.53 & -0.077 & -0.195 & 0.009 & 161 & 0.008 \\
F129--F150W & F090W--F150W & -0.38 & 0.07 & 0.009 & 0.201 & 0.009 & 51 & 0.012 \\
 &  & 0.07 & 1.64 & -0.001 & 0.334 & 0.012 & 575 & 0.012 \\
 & F115W--F150W & -0.13 & 0.25 & 0.000 & 0.546 & 0.004 & 175 & 0.007 \\
 &  & 0.25 & 0.96 & -0.016 & 0.611 & 0.007 & 599 & 0.007 \\
 & F150W--F200W & -0.17 & -0.03 & -0.001 & 0.449 & 0.012 & 27 & 0.027 \\
 &  & -0.03 & 0.07 & 0.033 & 1.525 & 0.028 & 153 & 0.027 \\
 &  & 0.07 & 0.22 & 0.068 & 1.047 & 0.025 & 334 & 0.027 \\
 &  & 0.22 & 0.57 & 0.153 & 0.653 & 0.031 & 267 & 0.027 \\
F158--F150W & F090W--F150W & -0.38 & 0.41 & 0.003 & -0.051 & 0.004 & 144 & 0.005 \\
 &  & 0.41 & 1.64 & 0.024 & -0.102 & 0.005 & 482 & 0.005 \\
 & F115W--F150W & -0.13 & 0.38 & 0.007 & -0.123 & 0.005 & 351 & 0.005 \\
 &  & 0.38 & 0.96 & 0.040 & -0.209 & 0.006 & 423 & 0.005 \\
 & F150W--F200W & -0.17 & 0.57 & -0.005 & -0.266 & 0.007 & 815 & 0.007 \\
F184--F200W & F090W--F200W & -0.51 & 2.14 & -0.008 & 0.019 & 0.009 & 623 & 0.009 \\
 & F115W--F200W & -0.27 & 0.10 & -0.007 & 0.086 & 0.007 & 72 & 0.008 \\
 &  & 0.10 & 0.85 & -0.001 & 0.016 & 0.008 & 554 & 0.008 \\
 &  & 0.85 & 1.53 & -0.047 & 0.070 & 0.009 & 144 & 0.008 \\
 & F150W--F200W & -0.17 & 0.06 & -0.006 & 0.179 & 0.006 & 153 & 0.008 \\
 &  & 0.06 & 0.28 & 0.003 & 0.035 & 0.008 & 519 & 0.008 \\
 &  & 0.28 & 0.57 & -0.029 & 0.151 & 0.009 & 142 & 0.008 \\
F213--F200W & F090W--F200W & -0.51 & 0.38 & 0.010 & -0.033 & 0.011 & 122 & 0.015 \\
 &  & 0.38 & 2.14 & -0.001 & -0.005 & 0.016 & 502 & 0.015 \\
 & F115W--F200W & -0.27 & 0.30 & 0.010 & -0.061 & 0.012 & 161 & 0.015 \\
 &  & 0.30 & 0.83 & -0.010 & 0.007 & 0.015 & 449 & 0.015 \\
 &  & 0.83 & 1.53 & 0.042 & -0.056 & 0.016 & 161 & 0.015 \\
 & F150W--F200W & -0.17 & 0.07 & 0.004 & -0.184 & 0.014 & 167 & 0.015 \\
 &  & 0.07 & 0.27 & -0.009 & 0.021 & 0.015 & 487 & 0.015 \\
 &  & 0.27 & 0.57 & 0.027 & -0.116 & 0.016 & 161 & 0.015 \\ \hline
\end{longtable*}

\subsection{Predictions for the tip of the red giant branch}

\begin{table*}[ht]
    \centering
    \begin{tabular}{RRR|RRR|RRRR|RRRR}
\hline \hline
\multicolumn{3}{c|}{Stellar parameters$^a$} & \multicolumn{3}{c|}{\Euclid} & \multicolumn{4}{c|}{\JWST} & \multicolumn{4}{c}{\Roman} \\
\mathrm{[Fe/H]} & \log L/L_{\odot} & \log T_{\mathrm{eff}} & Y_E & J_E & H_E & \mathrm{F090W} & \mathrm{F115W} & \mathrm{F150W} & \mathrm{F200W} & \mathrm{F087} & \mathrm{F129} & \mathrm{F158} & \mathrm{F213}\\ \hline
-3.20 & 3.180 & 3.651 & -4.38 & -4.89 & -5.36 & -4.10 & -4.53 & -5.13 & -5.40 & -4.04 & -4.77 & -5.21 & -5.41 \\
-2.50 & 3.239 & 3.646 & -4.54 & -5.06 & -5.55 & -4.25 & -4.68 & -5.30 & -5.58 & -4.19 & -4.93 & -5.40 & -5.59 \\
-2.20 & 3.264 & 3.637 & -4.60 & -5.14 & -5.67 & -4.31 & -4.76 & -5.40 & -5.70 & -4.25 & -5.02 & -5.50 & -5.71 \\
-1.90 & 3.288 & 3.623 & -4.66 & -5.24 & -5.83 & -4.36 & -4.83 & -5.53 & -5.85 & -4.29 & -5.10 & -5.64 & -5.86 \\
-1.70 & 3.305 & 3.612 & -4.70 & -5.30 & -5.93 & -4.39 & -4.88 & -5.61 & -5.96 & -4.31 & -5.16 & -5.73 & -5.97 \\
-1.55 & 3.317 & 3.603 & -4.73 & -5.35 & -6.03 & -4.41 & -4.91 & -5.68 & -6.05 & -4.33 & -5.21 & -5.81 & -6.06 \\
-1.40 & 3.329 & 3.592 & -4.75 & -5.40 & -6.11 & -4.42 & -4.94 & -5.75 & -6.14 & -4.32 & -5.25 & -5.88 & -6.16 \\
-1.30 & 3.337 & 3.585 & -4.76 & -5.44 & -6.17 & -4.43 & -4.96 & -5.80 & -6.21 & -4.33 & -5.28 & -5.93 & -6.23 \\
-1.20 & 3.345 & 3.577 & -4.78 & -5.48 & -6.24 & -4.43 & -4.99 & -5.85 & -6.28 & -4.33 & -5.32 & -5.99 & -6.30 \\
-1.05 & 3.357 & 3.566 & -4.81 & -5.54 & -6.32 & -4.44 & -5.03 & -5.91 & -6.38 & -4.33 & -5.37 & -6.06 & -6.40 \\
-0.90 & 3.369 & 3.555 & -4.85 & -5.59 & -6.40 & -4.44 & -5.07 & -5.98 & -6.47 & -4.31 & -5.42 & -6.13 & -6.49 \\
-0.70 & 3.385 & 3.537 & -4.91 & -5.68 & -6.53 & -4.42 & -5.14 & -6.09 & -6.61 & -4.28 & -5.51 & -6.25 & -6.64 \\
-0.60 & 3.392 & 3.528 & -4.94 & -5.73 & -6.59 & -4.41 & -5.18 & -6.14 & -6.68 & -4.26 & -5.55 & -6.30 & -6.71 \\
-0.40 & 3.404 & 3.511 & -5.01 & -5.83 & -6.70 & -4.36 & -5.26 & -6.24 & -6.82 & -4.20 & -5.64 & -6.41 & -6.85 \\
-0.30 & 3.410 & 3.501 & -5.05 & -5.88 & -6.76 & -4.33 & -5.31 & -6.30 & -6.88 & -4.16 & -5.69 & -6.47 & -6.92 \\
-0.20 & 3.414 & 3.492 & -5.08 & -5.93 & -6.81 & -4.29 & -5.34 & -6.34 & -6.95 & -4.11 & -5.73 & -6.51 & -6.98 \\
-0.08 & 3.417 & 3.481 & -5.14 & -5.99 & -6.87 & -4.23 & -5.40 & -6.40 & -7.02 & -4.04 & -5.79 & -6.57 & -7.05 \\
0.06 & 3.418 & 3.468 & -5.18 & -6.04 & -6.92 & -4.15 & -5.45 & -6.45 & -7.10 & -3.94 & -5.85 & -6.62 & -7.13 \\ \hline
\multicolumn{14}{l}{$^a$Reproduced from \citet{2021ApJ...908..102P} Table 4.} \\
    \end{tabular}
    \caption{TRGB absolute magnitudes in \Euclid, \JWST, and \Roman bands for BaSTI $\alpha$-enhanced 12.5 Gyr isochrones as a function of metallicity, transformed from $IJHK$ magnitudes originally reported in  \citet{2021ApJ...908..102P}.
    All magnitudes are in the Vega system.
    } \label{tab:basti_trgb}
\end{table*}
\begin{table*}[ht]
    \centering
    \begin{tabular}{LLL}
\hline \hline
\multicolumn{3}{c}{\Euclid} \\
M_{Y_E}^{\mathrm{TRGB}} &= \left\{ \begin{array}{l} 
      -4.51 -6.07[(Y_E-H_E)-0.98] \\
      -4.51 -0.79[(Y_E-H_E)-0.98] \\
\end{array}\label{eq:ye} \right. 
& \begin{array}{l}
0.85 < Y_E-H_E < 0.98 \\
0.98 \leq Y_E-H_E < 1.62 \\
\end{array} \\
M_{J_E}^{\mathrm{TRGB}} &= \left\{ \begin{array}{l} 
      -5.02 -6.55[(Y_E-H_E)-0.98] \\
      -5.02 -1.23[(Y_E-H_E)-0.98] \\
\end{array}\label{eq:je} \right. 
& \begin{array}{l}
0.85 < Y_E-H_E < 0.98 \\
0.98 \leq Y_E-H_E < 1.62 \\
\end{array} \\
M_{H_E}^{\mathrm{TRGB}} &= \left\{ \begin{array}{l} 
      -5.48 -7.07[(Y_E-H_E)-0.98] \\
      -5.48 -1.79[(Y_E-H_E)-0.98] \\
\end{array}\label{eq:he} \right. 
& \begin{array}{l}
0.85 < Y_E-H_E < 0.98 \\
0.98 \leq Y_E-H_E < 1.62 \\
\end{array} \\
\multicolumn{3}{c}{\JWST} \\
M_{\mathrm{F115W}}^{\mathrm{TRGB}} &= \left\{ \begin{array}{l} 
      -4.80 -1.41[(\mathrm{F115W}-\mathrm{F200W})-0.91] + 25.78[(\mathrm{F115W}-\mathrm{F200W})-0.91]^2 \\
      -4.80 -0.61[(\mathrm{F115W}-\mathrm{F200W})-0.91] -0.45[(\mathrm{F115W}-\mathrm{F200W})-0.91]^2 \\
\end{array}\label{eq:f115w} \right. 
& \begin{array}{l}
0.73 < \mathrm{F115W}-\mathrm{F200W} < 0.91 \\
0.91 \leq \mathrm{F115W}-\mathrm{F200W} < 1.71 \\
\end{array} \\
M_{\mathrm{F200W}}^{\mathrm{TRGB}} &= \left\{ \begin{array}{l} 
      -5.71 -2.41[(\mathrm{F115W}-\mathrm{F200W})-0.91] + 25.78[(\mathrm{F115W}-\mathrm{F200W})-0.91]^2 \\
      -5.71 -1.61[(\mathrm{F115W}-\mathrm{F200W})-0.91] -0.45[(\mathrm{F115W}-\mathrm{F200W})-0.91]^2 \\
\end{array}\label{eq:f200w} \right. 
& \begin{array}{l}
0.73 < \mathrm{F115W}-\mathrm{F200W} < 0.91 \\
0.91 \leq \mathrm{F115W}-\mathrm{F200W} < 1.71 \\
\end{array} \\
\multicolumn{3}{c}{\Roman} \\
M_{\mathrm{F106}}^{\mathrm{TRGB}} &= \left\{ \begin{array}{l} 
      -4.47 -6.96[(\mathrm{F106}-\mathrm{F158})-0.88] \\
      -4.47 -0.82[(\mathrm{F106}-\mathrm{F158})-0.88] \\
\end{array}\label{eq:f106} \right. 
& \begin{array}{l}
0.77 < \mathrm{F106}-\mathrm{F158} < 0.88 \\
0.88 \leq \mathrm{F106}-\mathrm{F158} < 1.85 \\
\end{array} \\
M_{\mathrm{F158}}^{\mathrm{TRGB}} &= \left\{ \begin{array}{l} 
      -5.34 -7.96[(\mathrm{F106}-\mathrm{F158})-0.88] \\
      -5.34 -1.82[(\mathrm{F106}-\mathrm{F158})-0.88] \\
\end{array}\label{eq:f158} \right. 
& \begin{array}{l}
0.77 < \mathrm{F106}-\mathrm{F158} < 0.88 \\
0.88 \leq \mathrm{F106}-\mathrm{F158} < 1.85 \\
\end{array} \\
M_{\mathrm{F129}}^{\mathrm{TRGB}} &= \left\{ \begin{array}{l} 
      -5.04 -1.12[(\mathrm{F129}-\mathrm{F213})-0.67] + 44.95[(\mathrm{F129}-\mathrm{F213})-0.67]^2 \\
      -5.04 -0.94[(\mathrm{F129}-\mathrm{F213})-0.67] -0.60[(\mathrm{F129}-\mathrm{F213})-0.67]^2 \\
\end{array}\label{eq:f129} \right. 
& \begin{array}{l}
0.52 < \mathrm{F129}-\mathrm{F213} < 0.67 \\
0.67 \leq \mathrm{F129}-\mathrm{F213} < 1.35 \\
\end{array} \\
M_{\mathrm{F213}}^{\mathrm{TRGB}} &= \left\{ \begin{array}{l} 
      -5.71 -2.12[(\mathrm{F129}-\mathrm{F213})-0.67] + 44.95[(\mathrm{F129}-\mathrm{F213})-0.67]^2 \\
      -5.71 -1.94[(\mathrm{F129}-\mathrm{F213})-0.67] -0.60[(\mathrm{F129}-\mathrm{F213})-0.67]^2 \\
\end{array}\label{eq:f213} \right. 
& \begin{array}{l}
0.52 < \mathrm{F129}-\mathrm{F213} < 0.67 \\
0.67 \leq \mathrm{F129}-\mathrm{F213} < 1.35 \\
\end{array} \\
\hline
    \end{tabular}
    \caption{Predicted color-absolute magnitude relations for the TRGB in \Euclid, \JWST, and \Roman bands based on $J-K_S$ and F110W--F160W relations from \citet{2017A&A...606A..33S}.}
    \label{tab:serenelli_trgb}
\end{table*}

\begin{figure}[ht]
    \centering
    \includegraphics[width=\columnwidth]{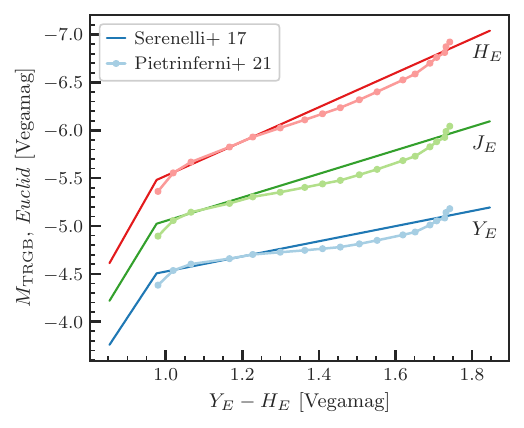}
    \caption{\Euclid/NISP $Y_E$, $J_E$, and $H_E$ TRGB magnitudes as a function of $Y_E - H_E$ color for the \citet{2017A&A...606A..33S} and \citet{2021ApJ...908..102P} theoretical TRGB calibrations.}
    \label{fig:euclid_trgb}
\end{figure}

\begin{figure}[ht]
    \centering
    \includegraphics[width=\columnwidth]{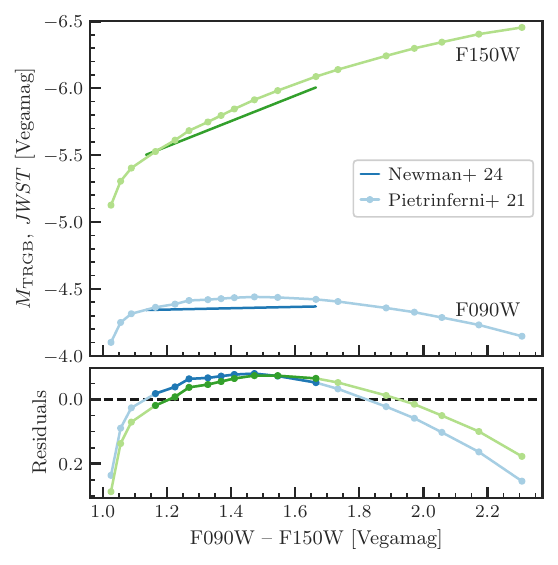}
    \caption{\JWST/NIRCam F090W and F150W TRGB magnitudes as a function of F090W -- F150W color for the \citet{2021ApJ...908..102P} theoretical and \citet{2024arXiv240603532N} empirical TRGB calibrations, with residuals in the lower panel.}
    \label{fig:jw_trgb}
\end{figure}

As a demonstration of how these transformations may be used, we convert theoretical predictions of the tip of the red giant branch absolute magnitude from ground-based and WFC3/IR bandpasses into the \JWST, \Roman, and \Euclid systems.
The tip of the red giant branch (TRGB) is the terminal point of a low-mass ($\lesssim$ 2 $M_{\odot}$) star's first ascent of the red giant branch, when core helium ignition halts the expansion and cooling of its atmosphere \citep{2006essp.book.....S}.
This produces a well-defined edge in color-magnitude space, which enables the TRGB to be used to great effect as a distance indicator in the local universe \citep[e.g.][]{1993ApJ...417..553L, 2018SSRv..214..113B, 2019ApJ...882...34F, 2024ApJ...966...89A}.
\JWST has already made TRGB detections out to 30 Mpc \citep{2024ApJ...961L..37C}.
With significantly smaller apertures, \Euclid and \Roman may not have the depth of \JWST, but they will deliver unprecedented spatial coverage of any and all Local Volume galaxies within their footprints, providing robust statistical samples of the ancient, metal-poor, dust-free halo populations where the TRGB is most effective as a distance indicator \citep{2018SSRv..214..113B, 2019ApJ...885..141B}.

We choose two recent theoretical calibrations of the TRGB to translate to \Euclid, \JWST, and \Roman bands.
While a plethora of empirical calibrations in the relevant bands exist as well \citep[e.g.][]{2004MNRAS.354..815V, 2012ApJS..198....6D, 2018AJ....156..278G, 2018ApJ...858...12H, 2019A&A...622A..63G, 2020ApJ...898...57D, 2024ApJ...966..175N}, we prefer these theoretical predictions for their larger dynamic ranges in color (metallicity).
\citet{2017A&A...606A..33S} establish a set of concordance physics for the TRGB with two stellar model suites and predict its color-absolute magnitude relations in the $VI$, $JK_S$, and F110W, F160W bands.
\citet{2021ApJ...908..102P} provide absolute $IJHK$ magnitudes for a 12.5 Gyr $\alpha$-enhanced TRGB as a function of metallicity with the BaSTI model suite.

We present transformed magnitudes for the BaSTI TRGB in \autoref{tab:basti_trgb}, and transformed color-magnitude relations from \citet{2017A&A...606A..33S} in \autoref{tab:serenelli_trgb}.
For \Euclid we use $Y_E - H_E$ as the color baseline throughout, as all \Euclid pointings will include all three NISP filters.
For \JWST and \Roman we convert the \citet{2017A&A...606A..33S} $JK_S$ relations to F115W, F200W and F129, F213 respectively, and the F110W, F160W relations to F106, F158.

In \autoref{fig:jw_trgb} we compare the BaSTI F090W, F150W TRGB to the recent empirical calibration of \citet{2024arXiv240603532N}, which we convert from Sirius-Vega to Vega magnitudes using the coefficients in \autoref{tab:zp_conv}.
The mean residuals over the valid color range of the empirical relation (1.14 $<$ F090W -- F150W $<$ 1.67 in Vega mags, corresponding to --1.9 $\le$ [Fe/H] $\le$ --0.7 in the BaSTI models) are --0.06 and --0.05 mag for F090W and F150W respectively, in the theoretical -- empirical direction.

\section{Discussion} \label{sec:disc}

The majority of our filter transformations require two or three components (49 and 28 filter combinations respectively), and the remainder require one or four (17 and 11).
As in \citet{2023AJ....166..236D}, the most consistent sources of slope changes are the Paschen jump, which occurs at 0.82~$\mu$m and peaks in A-type stars 
\citep{1986ApJS...60..577T, 1998BaltA...7..571S}, and the 1.65~$\mu$m bump, where the dominant source of H$^-$ continuum opacity transitions from bound-free to free-free absorption, most prominently in M stars \citep{1988A&A...193..189J, 2009ApJS..185..289R, 2023Atoms..11...61A}.

The most closely analogous filters (e.g., the synthetic magnitudes are within $\pm$0.1~mag across the full color range of our data) include: F090W and $z$; F106 and $Y_E$; F115W and F110W; F150W, F158, and F160W; and F200W, F213, and $K_S$.

\citet{2022A&A...662A..92E} provide transformations between NISP and several ground-based NIR instruments for models of O through K type stars, MLT stars, and galaxies.
Their $H_E - H_{\mathrm{2MASS}}$ vs.\ $J-H$ relation for O-K stars (their eqn.\ C.20) is the one most directly comparable to ours.
The slope of the reddest segment of ours, --0.107, is in good agreement with theirs, --0.12, which is dominated by redder stars.
However, our zeropoints differ by 0.05 mag when put onto the same magnitude system, and their color range is less than half of ours.

\citet{2023RNAAS...7..194S} and \citet{2024RNAAS...8..137S} derive absolute magnitude-spectral type relations for ultracool MLT-type dwarf stars in \JWST, \Roman, and \Euclid filters using the low-resolution IRTF/SpeX prism spectral libraries \citep{2014ASInC..11....7B}.
Although there is little if any overlap between the respective samples and spectral type coverage, when taken together these papers and ours are highly complementary, spanning $\sim$30 absolute magnitudes in the NIR.

\begin{figure}[ht]
    \centering
    \includegraphics[width=\columnwidth]{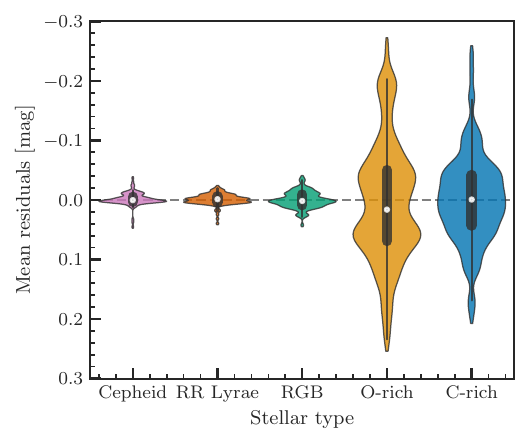}
    \caption{Distributions of average transformation residuals \added{(transformation-predicted magnitudes -- spectra-synthesized magnitudes)} for Cepheids, RR Lyrae, RGB stars, and O- and C-rich AGB stars.}
    \label{fig:residuals}
\end{figure}

In \autoref{fig:residuals} we show distributions of the average transformation residuals \added{(transformation-predicted magnitude -- spectra-synthesized magnitude)} for several subtypes of evolved stars, all of which are frequently used in distance scale work. 
For Cepheids, RR Lyrae, and RGB stars, the 16th and 84th percentile residuals are all within $\pm$0.01~mag, but for O-rich AGB stars they are --0.08, +0.10, and $\pm$0.06 for C-rich AGB stars.
Note that for RGB stars these residual dispersions may or may not directly translate to uncertainties on the TRGB, as we have few upper RGB stars and none at extreme low or high metallicities, where the TRGB slope is least certain \citep{2014AJ....148....7W,2017A&A...606A..33S}.
Similarly, the residual dispersions of the AGB stars do not necessarily indicate fundamental systematics on distance measurements made with them, only that cross-calibrating said measurements must be done in a more bespoke manner than with simple color-dependent transformations.

\section{Conclusions} \label{sec:conc}

We have presented transformations between NIR broadband filters on \JWST, \Roman, \Euclid, \HST, and several ground-based instruments.
We have additionally applied these transformations to predict color-magnitude relations for the theoretical tip of the red giant branch as seen with \Euclid and \JWST \added{(Figures~\ref{fig:euclid_trgb} and \ref{fig:jw_trgb} respectively, and \autoref{tab:serenelli_trgb})}, and \Roman \added{(\autoref{tab:serenelli_trgb})}.

In future, a comparison with state-of-the-art stellar atmosphere models such as the recently updated BOSZ library \citep{2017AJ....153..234B, 2024A&A...688A.197M} would be illuminating, both as a diagnostic of overall agreement between models and observations, and as a way to directly examine the effects of stellar physics and atmospheric composition---particularly carbon and other $\alpha$-element abundances---on broadband NIR fluxes.

A similar study of filter transforms for integrated light, such as with the XSL simple stellar population models \citep{2022A&A...661A..50V}, would be of interest as well for unresolved populations.

\begin{acknowledgments}
Support for this work was provided by NASA through HST program GO-15875. 
Support for this work was provided by the NSF through NSF grant AST-2108616.

This research has made use of the SVO Filter Profile Service (\url{http://svo2.cab.inta-csic.es/theory/fps/}) supported from the Spanish MINECO through grant AYA2017-84089.

This publication makes use of data products from the Two Micron All Sky Survey, which is a joint project of the University of Massachusetts and the Infrared Processing and Analysis Center/California Institute of Technology, funded by the National Aeronautics and Space Administration and the National Science Foundation.

This work has made use of data from the European Space Agency (ESA) mission \emph{Gaia} (\url{https://www.cosmos.esa.int/gaia}), processed by the \emph{Gaia} Data Processing and Analysis Consortium (DPAC, \url{https://www.cosmos.esa.int/web/gaia/dpac/consortium}). Funding for the DPAC has been provided by national institutions, in particular the institutions participating in the \emph{Gaia} Multilateral Agreement.

This research has made use of NASA’s Astrophysics Data System.

This research has made use of the SIMBAD database, operated at CDS, Strasbourg, France. The original description of the SIMBAD service was published in \citet{simbad_2000}.

This research has made use of the VizieR catalogue access tool, CDS, Strasbourg, France (DOI: 10.26093/cds/vizier). The original description of the VizieR service was published in \citet{vizier2000}.

This research made use of the cross-match service provided by CDS, Strasbourg.

This research has used data, tools or materials developed as part of the EXPLORE project that has received funding from the European Union’s Horizon 2020 research and innovation programme under grant agreement No 101004214.

\end{acknowledgments}

\facilities{CDS, MAST, IRTF, XShooter, HST:STIS, HST:NICMOS, HST:WFC3/IR}

\software{Astropy \citep{2013A&A...558A..33A, 2018AJ....156..123A},
          Astroquery \citep{2017ascl.soft08004G, 2019AJ....157...98G},
          Dust\_extinction \citep{dust_extinction},
          Dustmaps \citep{2018JOSS....3..695M},
          Matplotlib \citep{2007CSE.....9...90H},
          NumPy \citep{numpy, 2020Natur.585..357H},
          Pandas \citep{pandas, mckinney2011},
          pwlf \citep{pwlf},
          PyVO \citep{2014ascl.soft02004G},
          Seaborn \citep{2021JOSS....6.3021W},
          SciPy \citep{2020NatMe..17..261V},
          Specutils \citep{specutils},
          Stsynphot \citep{2020ascl.soft10003S},
          Synphot \citep{2018ascl.soft11001S}
}

\bibliography{bib,tmp,software}
\bibliographystyle{aasjournal}
\begin{appendix}

\section{Figures} \label{app:fit_figs}
\begin{figure*}[ht]
    \centering
    \includegraphics[width=\textwidth, height=0.9\textheight, keepaspectratio]{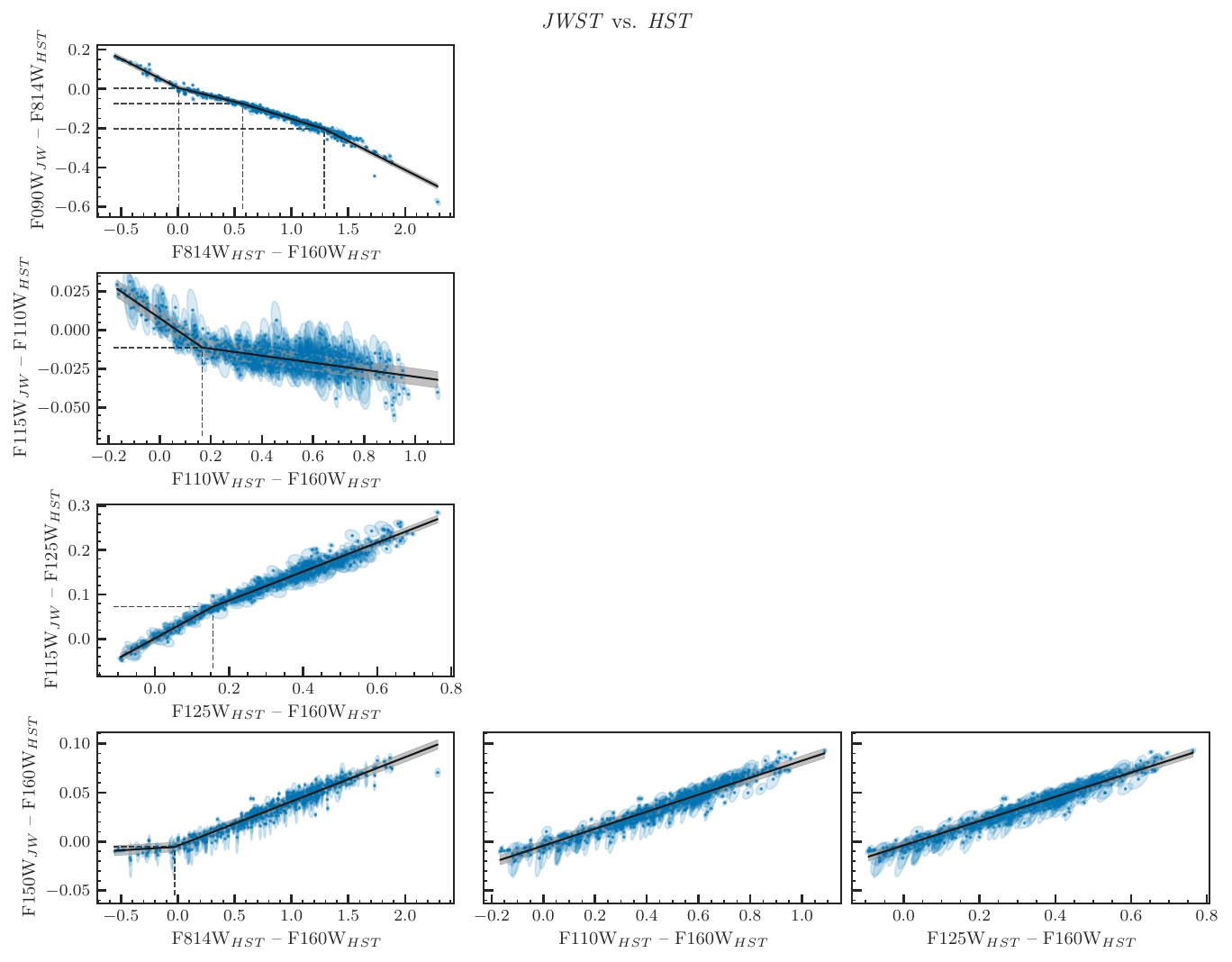}
    \caption{Filter transformations (black lines) for converting HST filters to JWST, with scatter as grey bands, and original data points and errors as blue points and ellipses.}
    \label{fig:jwst_hst}
\end{figure*}

\begin{figure*}[ht]
    \centering
    \includegraphics[width=\textwidth, height=0.9\textheight, keepaspectratio]{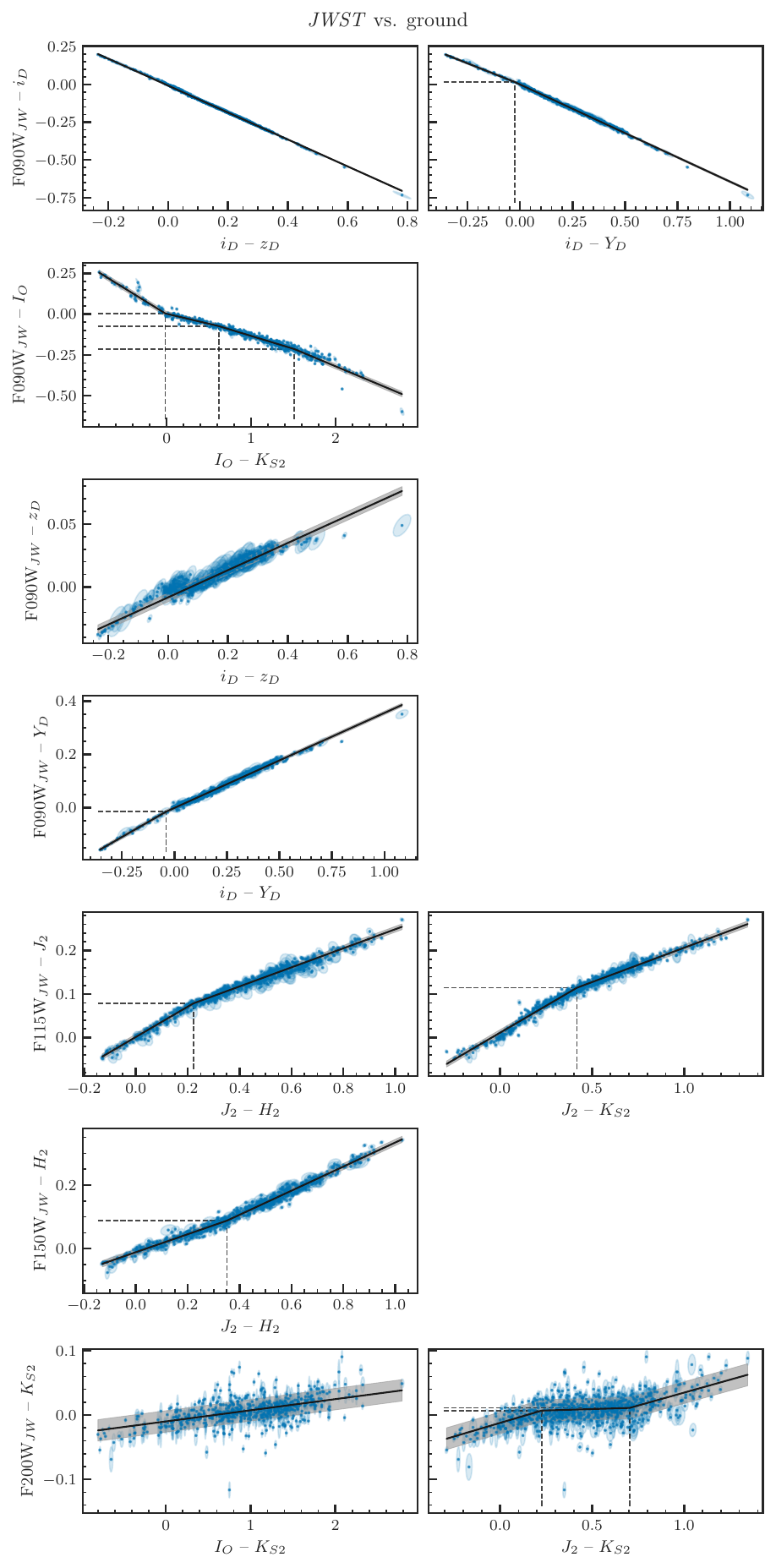}
    \caption{As \autoref{fig:jwst_hst} for converting ground filters to JWST.}
    \label{fig:jwst_ground}
\end{figure*}

\begin{figure*}[ht]
    \centering
    \includegraphics[width=\textwidth, height=0.9\textheight, keepaspectratio]{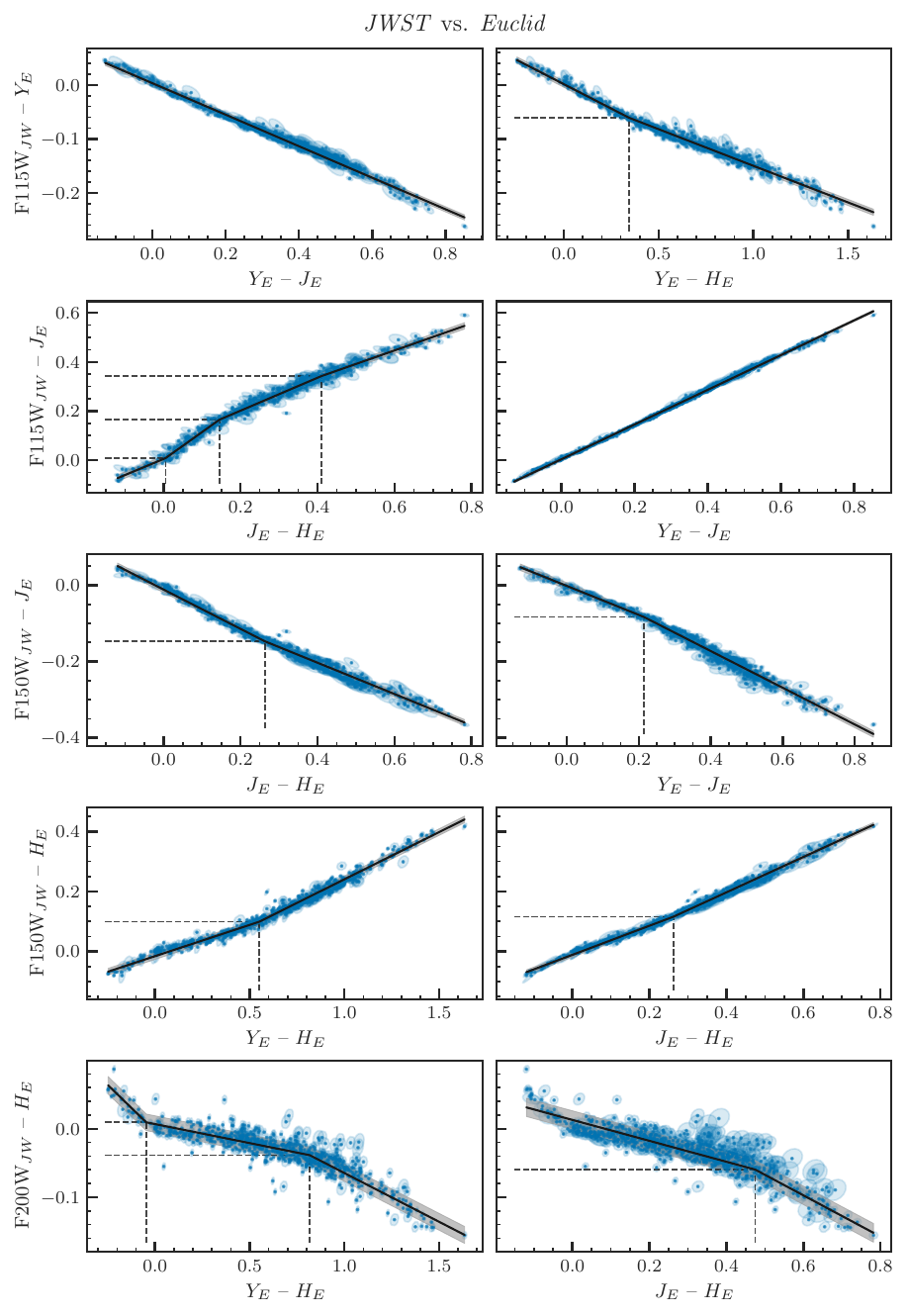}
    \caption{As \autoref{fig:jwst_hst} for converting Euclid filters to JWST.}
    \label{fig:jwst_euclid}
\end{figure*}

\begin{figure*}[ht]
    \centering
    \includegraphics[width=\textwidth, height=0.9\textheight, keepaspectratio]{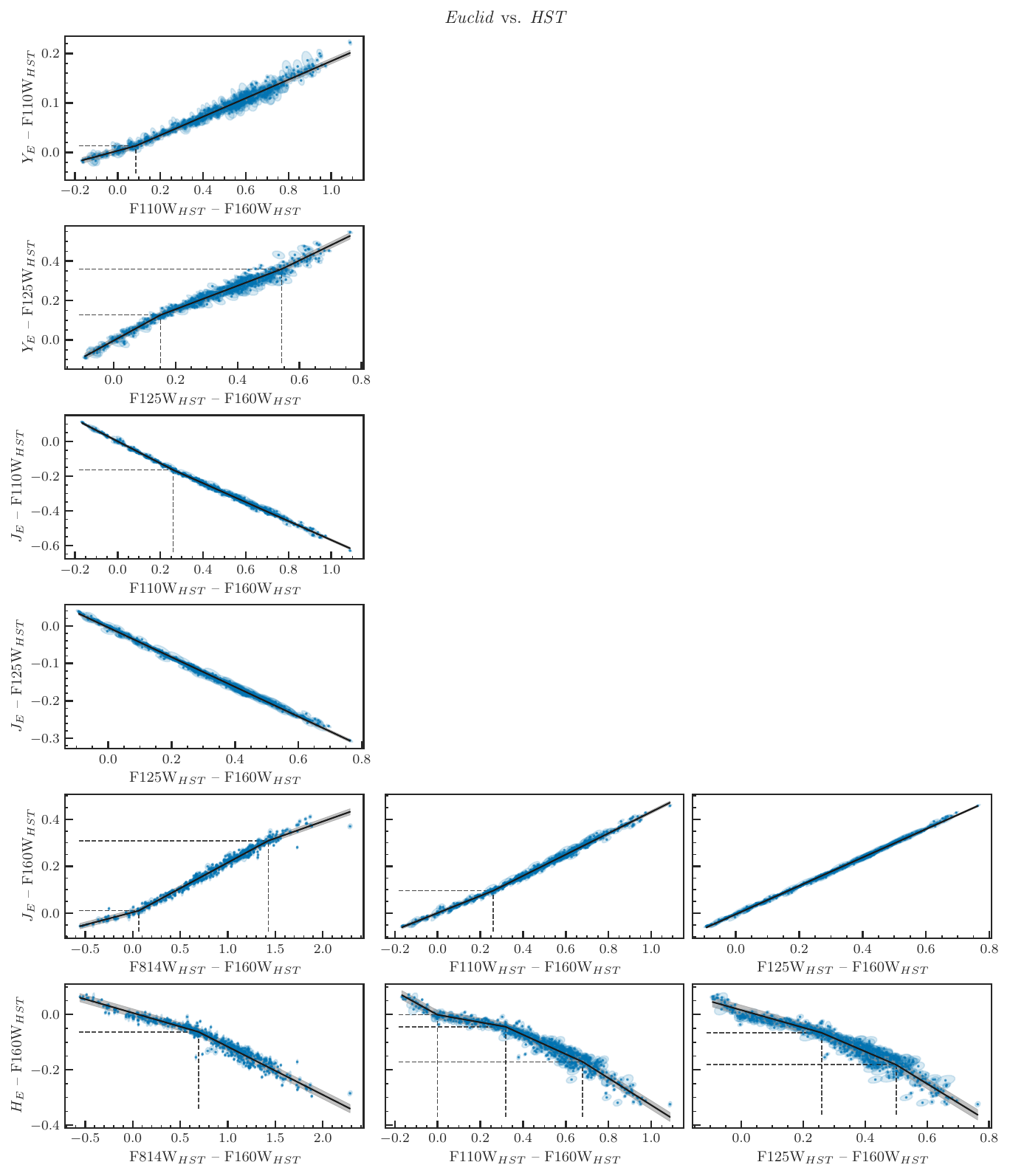}
    \caption{As \autoref{fig:jwst_hst} for converting HST filters to Euclid.}
    \label{fig:euclid_hst}
\end{figure*}

\begin{figure*}[ht]
    \centering
    \includegraphics[width=\textwidth, height=0.9\textheight, keepaspectratio]{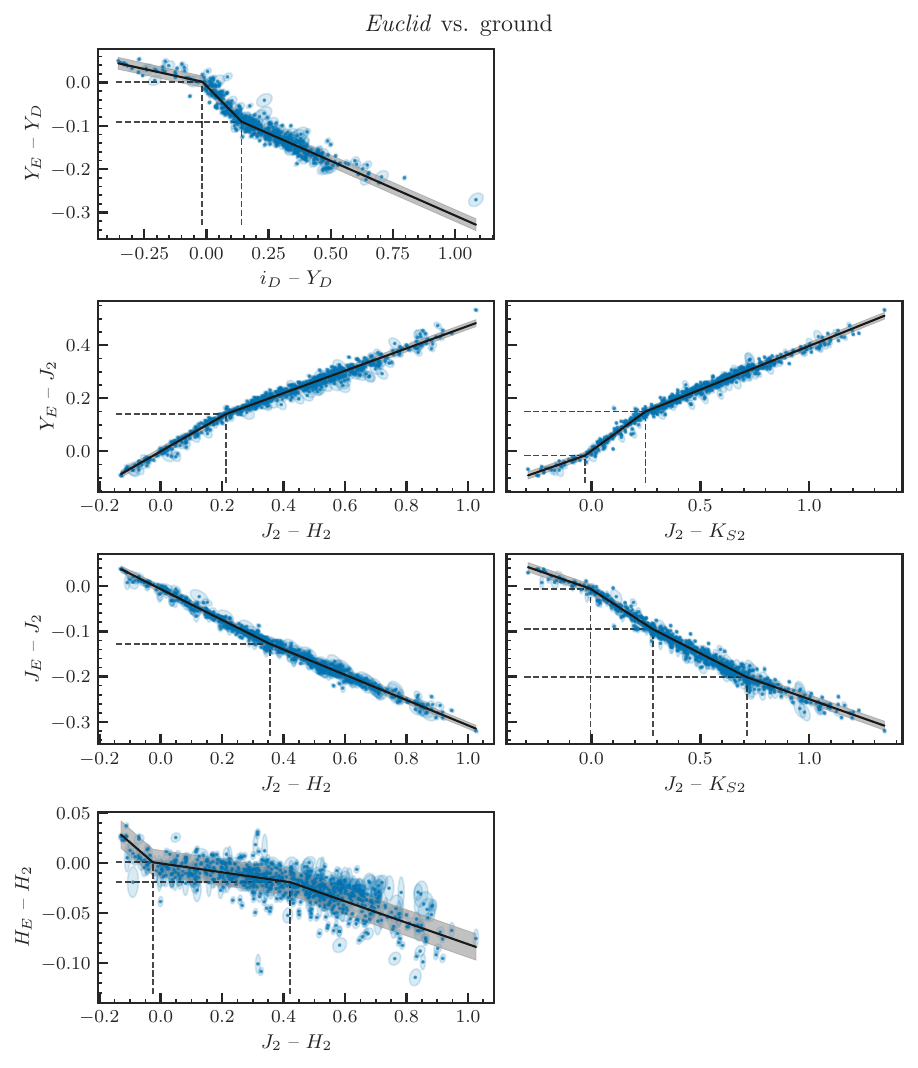}
    \caption{As \autoref{fig:jwst_hst} for converting ground filters to Euclid.}
    \label{fig:euclid_ground}
\end{figure*}

\begin{figure*}[ht]
    \centering
    \includegraphics[width=\textwidth, height=0.9\textheight, keepaspectratio]{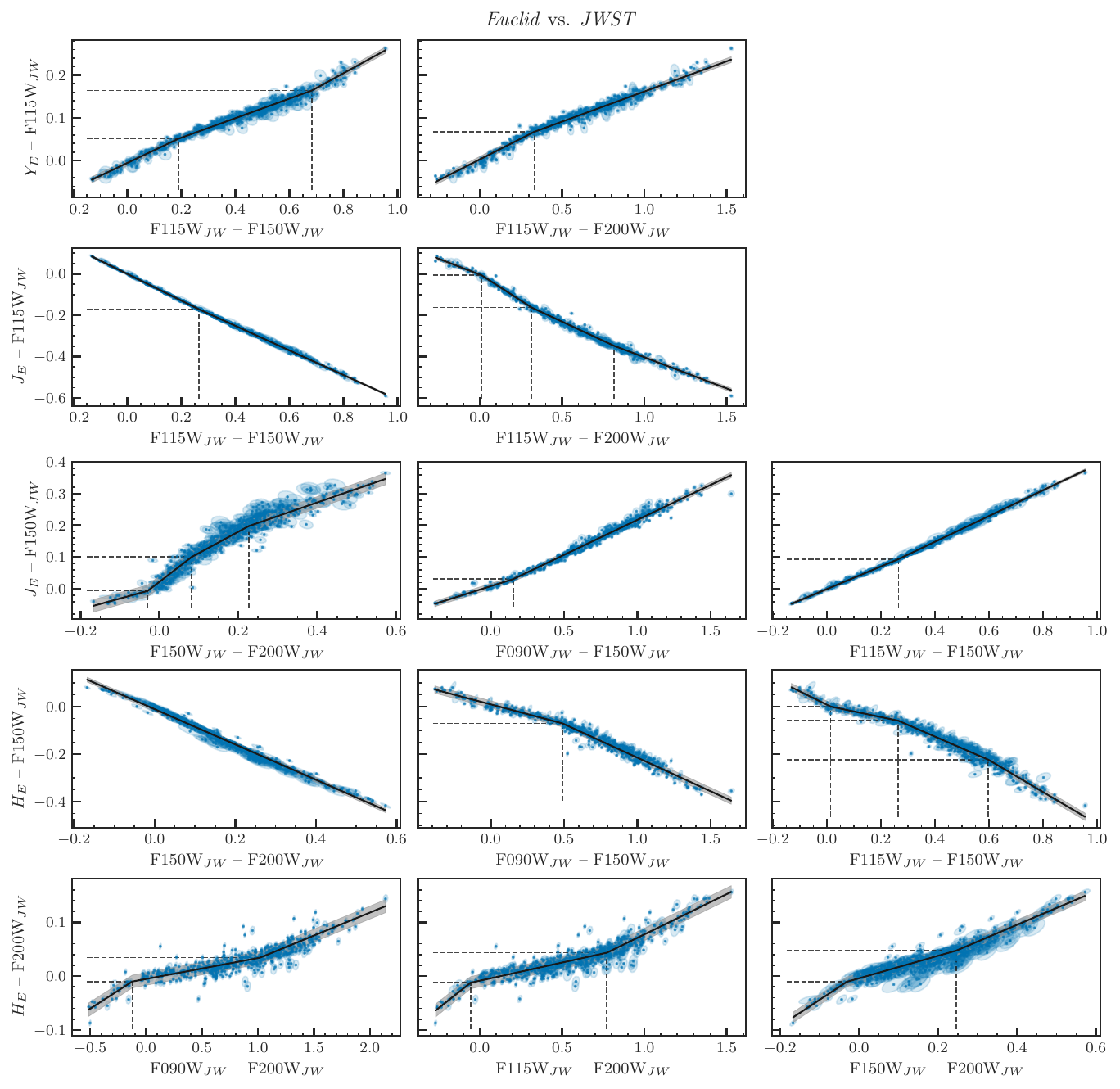}
    \caption{As \autoref{fig:jwst_hst} for converting JWST filters to Euclid.}
    \label{fig:euclid_jwst}
\end{figure*}

\begin{figure*}[ht]
    \centering
    \includegraphics[width=\textwidth, height=0.9\textheight, keepaspectratio]{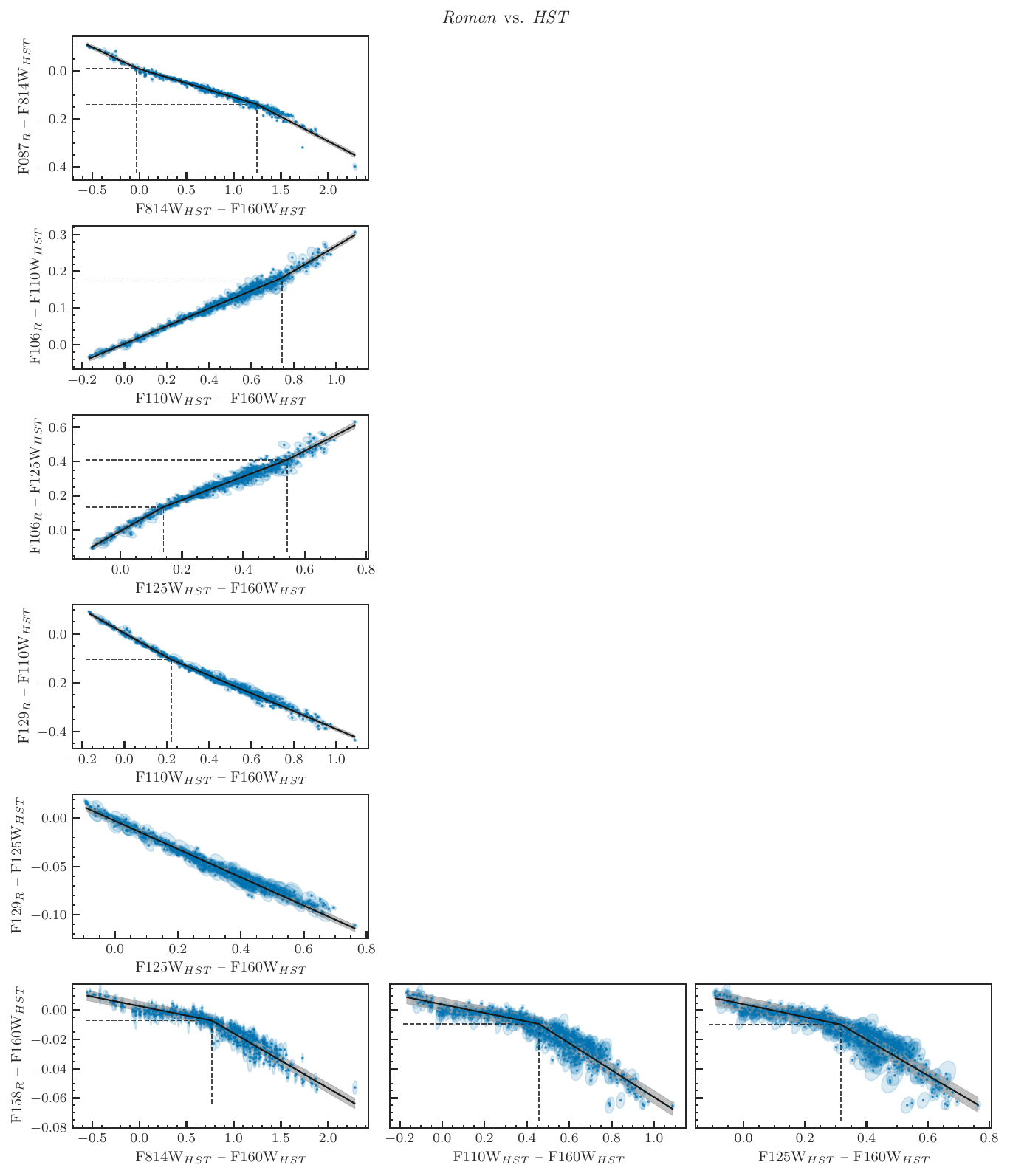}
    \caption{As \autoref{fig:jwst_hst} for converting HST filters to Roman.}
    \label{fig:roman_hst}
\end{figure*}

\begin{figure*}[ht]
    \centering
    \includegraphics[width=\textwidth, height=0.9\textheight, keepaspectratio]{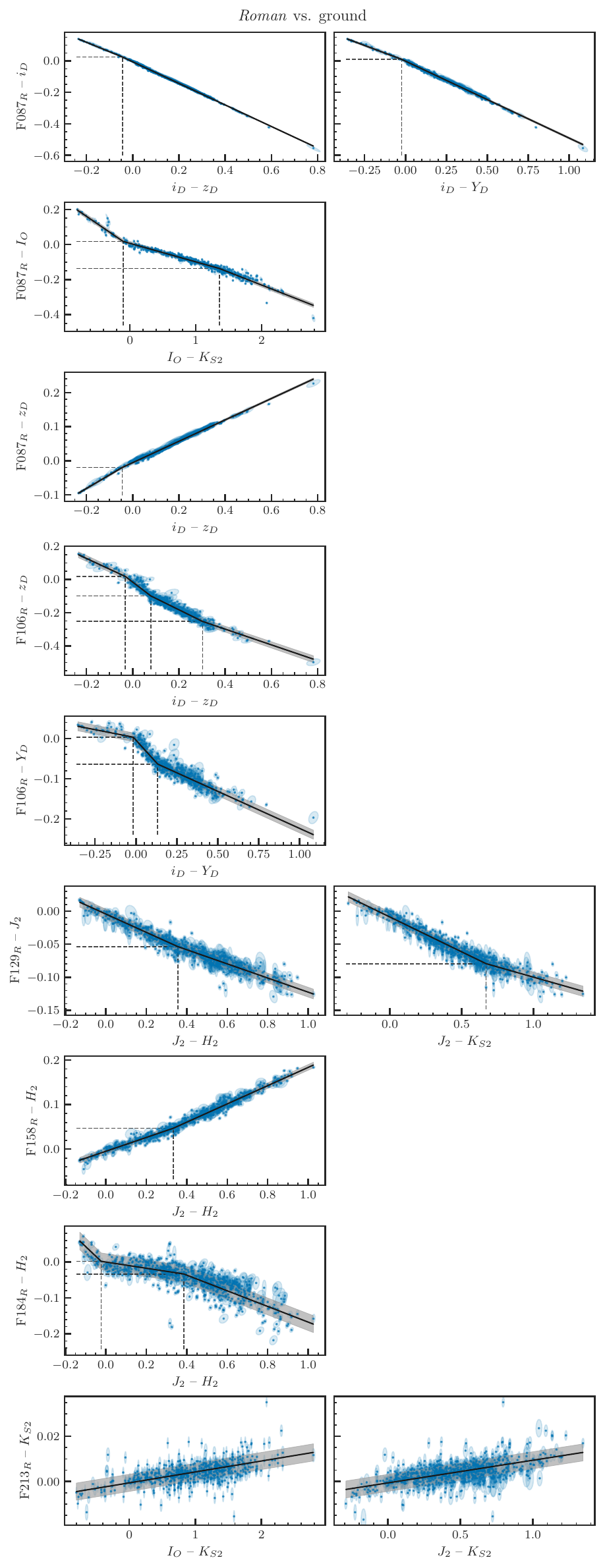}
    \caption{As \autoref{fig:jwst_hst} for converting ground filters to Roman.}
    \label{fig:roman_ground}
\end{figure*}

\begin{figure*}[ht]
    \centering
    \includegraphics[width=\textwidth, height=0.9\textheight, keepaspectratio]{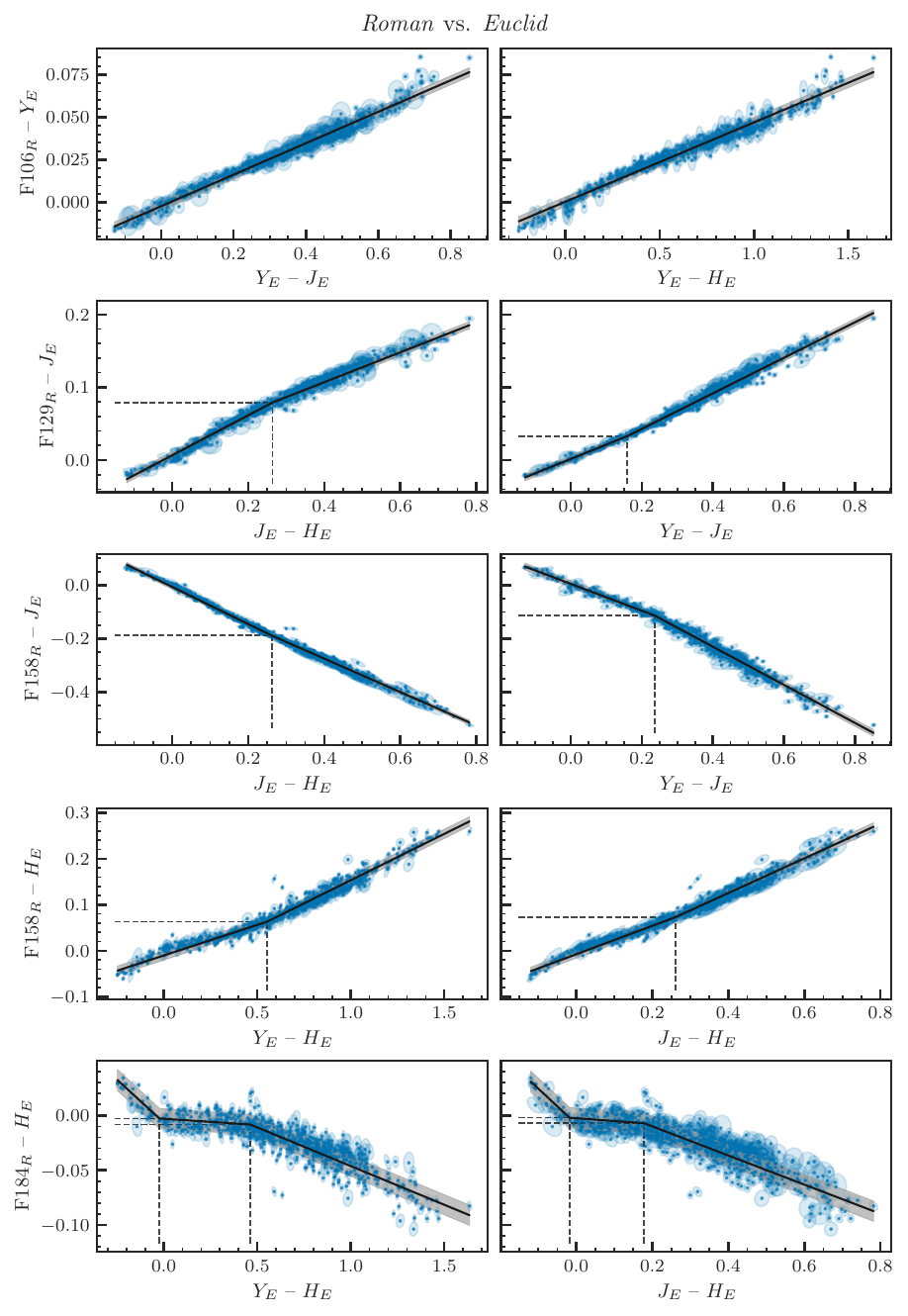}
    \caption{As \autoref{fig:jwst_hst} for converting Euclid filters to Roman.}
    \label{fig:roman_euclid}
\end{figure*}

\begin{figure*}[ht]
    \centering
    \includegraphics[width=\textwidth, height=0.9\textheight, keepaspectratio]{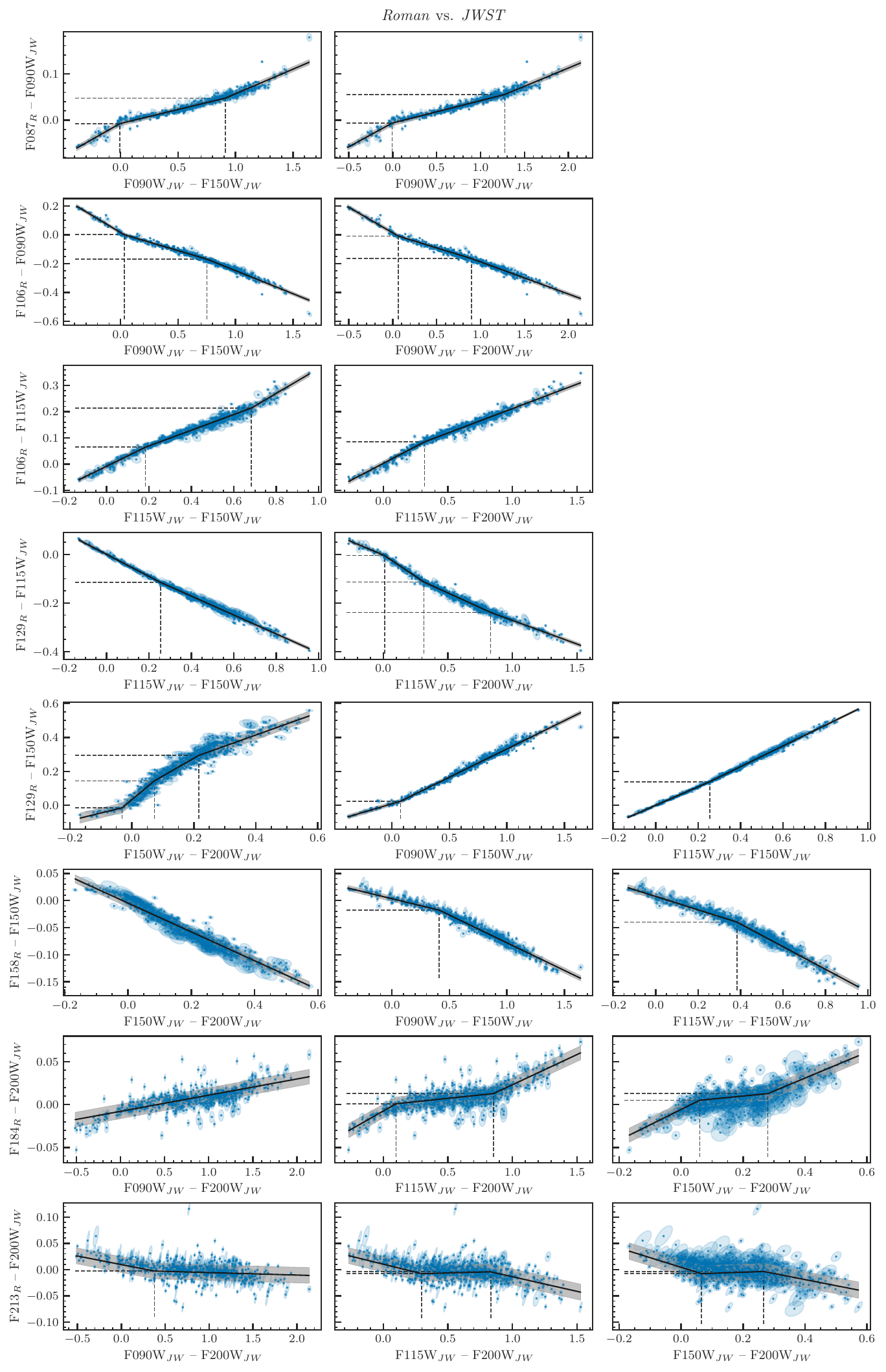}
    \caption{As \autoref{fig:jwst_hst} for converting JWST filters to Roman.}
    \label{fig:roman_jwst}
\end{figure*}

Figures~\ref{fig:jwst_hst} through \ref{fig:jwst_euclid} show \JWST vs.\ \HST, DECam $izY$ and 2MASS $JHK_S$, and Euclid respectively.
Figures~\ref{fig:euclid_hst} through \ref{fig:euclid_jwst} show Euclid vs.\ \HST, DECam+2MASS, and \JWST respectively.
Figures~\ref{fig:roman_hst} through \ref{fig:roman_jwst} show Roman vs.\ \HST, DECam+2MASS, Euclid, and \JWST respectively.

\end{appendix}

\end{document}